\documentclass[aps,prc,twocolumn,floatfix,showpacs,a4paper,
nofootinbib,amsmath,amssymb]{revtex4-1}
\usepackage{graphicx}
\usepackage{dcolumn}
\usepackage{bm}
\usepackage{color}

\newcommand{\be}{\begin{equation}}
\newcommand{\ee}{\end{equation}}
\newcommand{\ba}{\begin{eqnarray}}
\newcommand{\ea}{\end{eqnarray}}
\newcommand{\bd}{\begin{displaymath}}
\newcommand{\ed}{\end{displaymath}}
\renewcommand{\vec}[1]{\mbox{\boldmath$#1$}}

\begin{document}
\title{ Differential HBT Method to Analyze Rotation}

\author{L.P. Csernai and S. Velle}

\affiliation{
Institute of Physics and Technology, University of Bergen, 
Allegaten 55, 5007 Bergen, Norway 
}

\begin{abstract}
Two particle correlations are studied in the reaction plane of
peripheral relativistic heavy ion reactions where the initial
state has substantial angular momentum. The earlier predicted
rotation effect and Kelvin Helmholtz Instability, leads to
space-time momentum correlations among the emitted particles.
A specific combination of two particle correlation measurements
is proposed, which can sensitively detect the rotation of
the emitting system.  
Here the method is presented in simple 
 few source models where the symmetries and 
the possibilities of the detection can be demonstrated in a
transparent way.
\end{abstract}

\date{\today}


\maketitle

%%%%%%%%%%%%%%%%%%%%%%%%%%%%%%%%%%%
\section{Introduction}
%%%%%%%%%%%%%%%%%%%%%%%%%%%%%%%%%%%
\label{I}

Collective flow is one of the most dominant observable features
in heavy ion reactions up to the highest available energies, and
its global symmetries as well as its fluctuations are extensively
studied. Especially at the highest energies for peripheral reaction
the angular momentum of the initial state is substantial, which leads
to observable rotation according to fluid dynamical estimates
\cite{hydro1}.
Furthermore the low viscosity quark-gluon fluid may lead to
to initial turbulent instabilities, like the 
Kelvin Helmholtz Instability (KHI), according to numerical
fluid dynamical estimates
\cite{hydro2}, 
which is also confirmed in a simplified analytic model
\cite{KHI-Wang}. These turbulent phenomena further increase
the rotation of the system, which also leads to a large 
vorticity and circulation of the participant zone.
\cite{CMW12}.
In ref. \cite{hydro2} it is estimated that the increased rotation 
can be observable via the increased $v_1$-flow, but the $v_1$
signal at high energies is weak, so other observables of the
rotation are also needed.

The two particle correlation method is used to determine the
space-time size of the system emitting the observed particles,
thus providing valuable information on the exploding and expanding
system at the freeze out stage of a heavy ion collision. This method
is based on the Hanbury Brown and Twiss (HBT) method, originally used 
for the determination of the size of distant stars 
\cite{StarHBT}.
In heavy ion collisions the HBT method was used first for the
same purpose, the determination of the system size 
\cite{FirstHBTs},
but later also the ellipsoidal shape of the system and its tilt
\cite{DM95,MLisa}.
It was also observed relatively early that the expansion of the
system modifies the size estimates due to the collective radial
flow velocity of the emitting system
\cite{Pratt}, while the effect of flow on two particle correlations
was also analysed in great detail
\cite{Sin89}.
Transport model studies have indicated that the HBT radius
shows a minimum at the phase transition threshold
\cite{QfLi07-9}.

A detailed two particle correlation study of flow rotation
was not performed up to now \cite{LPSW}. This became actual as at
higher beam energies, where the initial angular momentum
of the participant system is increasing in peripheral reactions,
the system may rotate, causing a significant and detectable
effect. Here we do not want to use the most advanced state of the
art developments of the HBT method, rather we want to show in the
most simple way, how rotation can be detected by the method.

Different theoretical approaches were worked out up to
now to evaluate the two particle correlation functions
in different reaction models. We here picked one method,
which is not used very frequently up to now, but it can be 
generalized and used to fluid dynamical models with arbitrary
flow patterns well. With this method hereby we study
simplified, idealized fluid dynamical systems with different
symmetry structures. These studies show how we can detect
rotation via two particle correlation functions and what
effects may cause difficulties in identifying rotation.

Based on these studies we also present a Differential 
Hanbury Brown and Twiss (DHBT) method, 
which can sensitively determine the strength
(vorticity or circulation) of the rotating flow and
the direction of this rotation.

%%%%%%%%%%%%%%%%%%%%%%%%%%%%%%%%%%%
\subsection{The Emission Function}
%%%%%%%%%%%%%%%%%%%%%%%%%%%%%%%%%%%
\label{TEF}

Following \cite{Cs-0} the definition of a particle 4-current is: 

\be
N^\mu = \int p^\mu \frac{d^3p}{p^0} f(x,p) \ ,
\ee
where $f(x,p)$ is the invariant scalar phase space density distribution 
of the emitted particles.
The total flow of N nucleons across a space-time (ST) hypersurface, with
the surface element, $d\sigma_\mu$ is:
\begin{equation}
N = \int \frac{d^3p}{p^0} \int d\sigma_\mu\, p^\mu f(x,p) \ .
\end{equation}
If we do not perform the integration over the momentum, $p$, then we get
the Cooper-Frye formula \cite{CF} for invariant momentum distribution:
\begin{equation}
E\frac{dN}{d^3 p} = \int d\sigma_\mu\, p^\mu f(x,p)\ ,
\end{equation}
where $E=p^0$.
This assumes that there is a 3-dimensional Freeze Out (FO) hypersurface, 
which can also be generalized to become a layer, as we will show.
Using $d^3\vec p = dp_{||}d \vec p_\perp = p^0 dy d \vec p_\perp )$
we can write this in another form, and we can extend it to a 4-volume
integral of a 4-dimensional Source Function, $S$, as  \cite{CYW, WF10}:
\begin{equation}
\frac{dN}{dyd^2p_\perp} = 
\int d\sigma_\mu (x) \, p^\mu f(x,p) = 
\int d^4 x\, S(x, \vec p) \ , 
\end{equation}
where we assumed that 
the emission appears on a 3-dimensional hypersurface with the 
outward pointing normal, $d\sigma^\mu$.
The emission function gives the distribution of the ST positions 
of momenta of emitted particles. The emission function gives the number 
of the particles, $\Delta N$, emitted in the phase-space element 
$ \Delta ^3 x \Delta ^3 p$ per unit time, $\Delta t$. 
We get a Lorentz invariant scalar if we also multiply it by the 
energy, $p^0$, of the emitted particles:
\begin{equation}
S(x,\vec p) = p^0 \frac{\Delta N}{\Delta t \Delta ^3 x \Delta ^3 p} \ .
\end{equation}
Along the lines of this introduction we can describe the emission over
a hypersurface also as a 4-volume integral.
We still assume a ST hypersurface with an outward pointing  
normal vector, $d\sigma^\mu$, but then the ST integral is interpreted 
as a 4-volume integral with a delta function for the given surface, 
$\delta (x'-x)$ as: 
\begin{equation}
\int S(x, \vec p )\, d^4x =
\int d^4x\,  p^\mu\, \hat\sigma_\mu(x')\, \delta^{(4)}(x'-x)\, f(x,p)\ ,
\end{equation}
where the emission is constrained to a 3-dimensional hypersurface
in the ST and it is directed in a given direction characterized by
the unit vector in the source described by 
$\hat\sigma_\mu(x')\, \delta^{(4)}(x')$.
Even if we assume that the source is in a ST layer (which is not 
too thick, e.g. 2-3 fm or 2-3 fm/c), in this layer we can 
have a maximum of the emission. This FO layer in case of pions is narrow
if we consider the rapid and simultaneous hadronization and FO from
the plasma.
This could even be idealized as a 3-dim hypersurface in the 4-dim ST
if the thickness of the layer is neglected. Thus, $\hat\sigma_\mu$, 
is the unit normal vector of this surface or layer can 
be both timelike or spacelike:
\ba
\hat\sigma^\mu \hat\sigma_\mu &=& +1 \ 
{\rm for\ timelike\ hypersurface\ or\ layer} \ ,
\nonumber\\
\hat\sigma^\mu \hat\sigma_\mu &=& -1 \ 
{\rm for\ spacelike\ hypersurface\ or\ layer} \ .
\ea

The idealization of FO in a  3D hypersurface is not necessary,
however it makes the presentation more transparent.  The more realistic 
emission distribution must happen in a 4D ST surface layer, which 
can still can have an
effective space-like or time-like normal vector. See 
refs.\,\cite{Sin89,M-2,M-3,M-4}. 
Although, one might naively believe that in case of a
normal vector, $\hat\sigma^\mu = (1,0,0,0)$ the emission is uniform in all
spatial directions, this is not true, as the local flow velocity
also influences the emission probability 
\cite{Sin89,M-3}. When the flow velocity points in the direction of the
detector (or the FO normal) the probability for the emission
into the $\hat\sigma^\mu$-direction is bigger, so that the 
emission probability should be proportional to $(u^\mu \cdot \hat\sigma_\mu)$.
This will be important later on when the observability of rotation
is discussed. 

At ultrarelativistic energies most of the FO happens in time like
directions, frequently idealized as $t=$constant or $\tau=$constant
hypersurfaces or layers.
In addition even in case of time-like FO the deeper (or earlier)
points of the FO layer have a smaller emission probability because
of the opacity of QGP, indicated also by the strong jet quenching. 
This effect causes additional asymmetries in the emission.

%%%%%%%%%%%%%%%%%%%%%%%%%%%%%%%%%%%
\subsection{Hydrodynamical Parameterization}
%%%%%%%%%%%%%%%%%%%%%%%%%%%%%%%%%%%
\label{HP}

Let us assume that at the points of the source the matter is still
in local equilibrium. Then we can describe the phase-space distribution 
of the particles, $f(x,p)$ by a J\"uttner distributions \cite{Cs-0} or 
a relativistic
Bose-Einstein or Fermi-Dirac distribution.
Furthermore, instead of the delta function we may assume the emission 
distributed in a ST layer, which still has a preferred direction of
emission $\hat\sigma^\mu$. 

For our purposes the most suitable parametrization of the emission 
function is introduced in the special "Buda-Lund" model, see 
section 8 of ref.
\cite{Cso-5} . 
Here the emission function is parametrized as
\begin{equation}
S(x, \vec p)\, d^4 x =
p^\mu\, d^4 \Sigma_\mu(x)\, f(x,p) ,
\label{S-BL}
\end{equation}
where the emission probability and its dependence of the FO direction
is already included in the $p^\mu\, d^4 \Sigma_\mu(x)$ term.
The 4-volume integral is directed and yields a maximum for $k^\mu$
which is closest to $d^4 \Sigma^\mu(x)$. In the Buda-Lund model the
FO direction points into the flow 4-velocity, 
$d^4 \Sigma^\mu(x) \propto u^\mu $. This is also a frequent approximation
in other fluid dynamical models, although, in the general case
it is not a valid approximation. Thus, this approach is identical to the
dynamical volume FO in a layer 
\cite{M-2,M-3,M-4},
discussed above. We also assume that the FO layer depends parametrically
on the proper time, $\tau$, (or distance $s$) in the
FO factor $d^4 \Sigma^\mu(x) \propto \hat\sigma^\mu d^4 x$. So, that the
emission probability is proportional to \ $G(x) \, H(\tau)$:
$$
S(x,\vec p) d^4 x = 
  p^\mu\  \hat\sigma_\mu(x) \ G(x)\, H(\tau)\, \ d\tau d^3x \  f(x,p) \ ,
$$
where 
\begin{equation}
H(\tau) = \frac{1}{(2\pi \Theta^2 )^{1/2}} 
\exp\left[-\frac{(\tau-\bar{\tau})^2}{2 \Theta^2 }\right] ,
\label{S-BL2}
\end{equation}
and that the widths of the emitting sources do not change significantly 
during the course of emission from a given source. (The emission probability
will be discussed in more detail when the source asymmetry
is analysed in section \ref{ASm}.)
The emission function characterized with a 
locally thermalized volume-emitting source is then:
\begin{equation}
S(x, \vec p)\, d^4 x =
\frac{g}{(2\pi\hbar)^3} 
\frac{p^\mu\, \hat\sigma_\mu(x)\, G(x)\, H(\tau)\, d\tau d^3x}
{\exp\left(\frac{p^\mu u_\mu (x)}{T(x)}-\frac{\mu (x)}{T(x)}\right)-1},
\label{S-hydro}
\end{equation}
where in place of $f(x,p)$ we inserted the relativistic Bose-Einstein 
distribution. The  factor $g$ stands for the degeneracy, $u^\mu(x)$ is 
the 4-velocity field, $T(x)$ is the temperature field, $\mu (x)$ is 
the chemical potential and $d^4 x = d\tau\,dx\,dy\,dz$.
$G(x)$ is the ST emission density across the layer of the particles 
(e.g. pions). This can be approximated with a ST hypersurface
and then with $\Theta \rightarrow 0$ we obtain the 
Cooper-Frye FO description.
\cite{CF}

For the phase space distribution we frequently use the 
J\"uttner (relativistic Boltzmann) distribution:
\be
f^J(x,p) = \frac{g}{(2\pi\hbar)^3} 
\exp\left(-\frac{p^\mu u_\mu (x)}{T(x)}+\frac{\mu (x)}{T(x)}\right),
\label{Jut-1}
\ee
which is normalized to the invariant scalar density of particles
\be
n(x) = N^\mu u_\mu = u_\mu \int \frac{d^3p}{p^0} p^\mu\, f(x,p) =
\frac{g e^{\mu/T}}{(2\pi\hbar)^3}C_n,
\ee
where $C_n = 4 \pi m^2 T K_2(m/T)$
and we use the $c = k = 1$ convention.
Thus in terms of the local invariant
scalar particle density the J\"uttner distribution is \cite{Cs-0}
\be
f^J(x,p) = \frac{n(x)}{C_n} 
\exp\left(-\frac{p^\mu u_\mu (x)}{T(x)}\right).
\label{Jut-2}
\ee

We can also define $d\sigma^\mu = dx\,dy\,dz\ \hat\sigma^\mu$ (for a 
timelike surface or layer) where the norm of $d\sigma^\mu$ is the 
3-volume of the source element (like a fluid cell), 
similarly to Refs. 
\cite{M-2,M-3,M-4}

Our source function in this case, with
Eq. (\ref{Jut-2}), similarly to Ref. \cite{Cso-5}
in the frame where $\hat\sigma^\mu = (1,0,0,0)$ is:
\begin{equation}
G(x) = N_{FD} (x) / n(x) ,
\label{G1}
\end{equation}

where $ N_{FD} (x)$ is the density of particles arising from 
fluid-dynamical (FD) evolution (or from other transport models). 
Then taking inti account the flow velocity in this frame:
\begin{equation}
N_{FD} (x) =  \gamma n (x) \ .
\end{equation}
The $H(\tau)$ (or $H(s)$) freeze-out probability along 
the $\tau$ (or $s$) parameter,
across the layer can be integrated separately from the remaining three
orthogonal coordinates.

Here we assume that the primary direction of the emission is
$\hat{\sigma}^\mu$. Thus, in case of an explosively expanding
system it points towards the detector, so 
$\hat{\sigma}^\mu \approx \hat{k}^\mu$.
The emission happens from the 4-volume of ST surface layer with 
an effective normal direction $\hat{\sigma}^\mu$, and not from a
ST hypersurface.

In ref. \cite{Sin89} the correlation function was analyzed in detail
in dependence of the direction of the primary emission direction,
$k^\mu$. A possibility of longitudinal momentum difference was 
considered in terms of the rapidity of the two emitted particles,
where the difference of the longitudinal momenta, the width parameter,
was studied in detail. While this analysis provides a deeper insight 
into the features of the correlation function, in our studies we 
concentrate to the rotation of the system and
restrict ourself to a simplest presentation of the correlation function,
which can be realized experimentally without much additional effort.  

We can assume FO from a narrow layer at a proper time hyperbola 
$\tau_{FO} = cons.$ like in the Buda-Lund model, see 
ref.\,\cite{Cso-5}. 
This can be practical if the CFD
model uses proper time and rapidity coordinates. Recent studies indicate
that irrespective of the coordinate system choice, in the major part
of FO in high energy collisions the FO happens near to a constant
proper time hyperbola, although the origin of this FO-hyperbola
is at an earlier point of time than the intersection of the
centers of the projectile and target trajectories 
\cite{Anchiskin}.

For the first test purpose we take an oversimplified model of 4 fluid 
elements, which may or may not expand or rotate.
We assume that these are in the reaction plane, $[x-z]$-plane, and
will characterise parameters of a heavy ion reaction based on CFD
results. We assume that the system is stationary so the time emission 
probability is a Gaussian (like) distribution in time. Later on we 
intend to study and see that these methods can be applied for realistic
full scale FD calculations also.

%%%%%%%%%%%%%%%%%%%%%%%%%%%%%%%%%%%
\subsection{Pion Correlation Functions}
%%%%%%%%%%%%%%%%%%%%%%%%%%%%%%%%%%%
\label{PCF}

The pion correlation function is defined as the inclusive two-particle 
distribution divided by the product of the inclusive one-particle 
distributions, such that \cite{WF10}:
\begin{equation}
C( p_1 , p_2) = 
\frac{P_2( p_1, p_2)}{P_1( p_1)P_1( p_2)},
\end{equation}
where $ p_1$ and $ p_2$ are the 4-momenta of the pions. 

We will assume pions are created at two points, $ x_1$ and $ x_2$,
which are distributed in space. The particle distribution is given by the
reduced phase space source distribution:
\be
S(x,{p}) =
f(x,{p})\, p^\mu\, \hat\sigma_\mu\, G(x)\,H(\tau) \ . 
\ee
For two identical pions with momenta 
$ p_1$ and $ p_2$ the two-particle distribution is:
\begin{equation}
P_2 ( p_1,  p_2) = \int d^4 x_1 d^4 x_2 
S(x_1, p_1) S(x_2, p_2) |\psi_{12} |^2,
\label{P2}
\end{equation}
where the wave equation $\psi_{12}$ is given by:
\begin{equation}
\psi_{12} = \frac{1}{\sqrt{2}} 
(e^{i p_1 \cdot  x_1 + i p_2 \cdot  x_2 }+
 e^{i p_1 \cdot  x_2 + i p_2 \cdot  x_1 }) \ .
\end{equation}
We now introduce
the center-of-mass momentum
\footnote{The vector $\vec k$ is the wavenumber vector,
$ \vec k = \vec p/\hbar$ so for numerical calculations we have to
use that $\hbar c =$ 197.327 MeV fm.}
\begin{equation}
 k = \frac{1}{2} ( p_1 +  p_2),
\end{equation}
and the relative momentum
\begin{equation}
 q =  p_1 -  p_2,
\end{equation}
where assuming the mass-shell constraint for the two
particles and so we have $q \cdot k = (p_1-p_2)\cdot(p_1+p_2)/2 = 
(p_1^2 - p_2^2)/2 = (m_\pi^2 - m_\pi^2)/2 = 0$, which leads to
$q^0 = \vec q \cdot \vec k / k^0$

With the relative and center-of-mass momentum we can write the 
wave equation as:
\begin{equation}
\psi_{12} = 
\frac{e^{i k \cdot ( x_1 + x_2)}}
     {\sqrt{2}} 
\left( e^{ i q \cdot ( x_1 - x_2)/2} + 
       e^{-i q \cdot ( x_1 - x_2)/2} \right)\ ,
\label{Psi12q}
\end{equation}
and then
\be
|\psi_{12}|^2 = 
\left[ 1 + \frac{1}{2}
\left(e^{ i q \cdot (x_1 - x_2)} + 
      e^{-i q \cdot (x_1 - x_2)}\right) \right] \ .
\label{Psi12k}
\ee
We can then insert Eq. (\ref{Psi12k}) into Eq. (\ref{P2}), and we obtain:
\ba
P_2 ( p_1,  p_2) &=& 
\int d^4 x_1\, d^4 x_2\, 
S(x_1,  {k}+{q}/2)\, S(x_2,{k}-{q}/2)
\nonumber \\
& \times & 
\left[ 1 + \frac{1}{2}
\left(e^{ i q \cdot (x_1 - x_2)} + 
      e^{-i q \cdot (x_1 - x_2)}\right) \right] \ ,
\label{P2b}
\ea
where the last term in the brackets is $\cos[q(x_1 - x_2)]$.
Similarly for the one particle distribution we get:
\begin{equation}
P_1( p) = 
\int d^4 x\ S(x, k) \ .
\label{P1b}
\end{equation}
We now use a method for
moving sources presented in ref.\,\cite{Sinyukov-1}. 
Using Eqs. (\ref{P2b},\ref{P1b}), together with the definition of the 
correlation function we have:
\begin{equation}
C(k,q) =
1 + \frac{R(k,q)}{\left| \int d^4 x\,  S(x, k) \right|^2}\ ,
\label{C-def}
\end{equation}
where
\be
\begin{split}
R(k,q) =&  \int d^4 x_1\, d^4 x_2\, \cos[q(x_1-x_2)] \times\\
& S(x_1,  {k}+{q}/2)\, S(x_2,{k}-{q}/2)\ .
\end{split}
\label{R1}
\ee
Here $R(k,q)$ can be calculated \cite{Sinyukov-1} via the function
\be
\begin{split}
& J(k,q) = \int d^4x\ S(x,k+q/2)\, \exp(iqx) = \\
        &  \int d^4x\ S(x,k+q/2)\, [\cos(qx) + i \sin(qx)] \ ,
\end{split}
\label{J-def}
\ee
and we obtain the $R(k,q)$ function as
\be
R(k,q) = {\rm Re}\, [ J(k,q)\, J(k,-q) ]
\label{R-def}
\ee

This can be verified, by
using Eq. (\ref{J-def}),
forming a double integral over $d^4x_1\, d^4x_2$ from 
$J(k,q)\, J(k,-q)$, yielding to a term $\exp[-iq(x_1-x_2)]$. Then
taking the real part of the double integral leads to a term
$\cos[q(x_1-x_2)]$ and this recovers Eq. (\ref{R1}).

\bigskip
%%%%%%%%%%%%%%%%%%%%%%%%%%
{\bf Source with local J\"utner distribution:}
Let us take the $S (x_1, p_1) S (x_2, p_2) $ term in Eq. (\ref{P2b}),
and assume that the single particle distributions, $f(x,p)$, in
the source functions are J\"uttner distributions, which depend on the
local velocity, $u^\mu(x)$ , via the term
\be
\exp\left(\frac{- p^\mu u_\mu (x)}{T(x)}\right)
\ee
as shown in Eq. (\ref{Jut-2}). Here the
local flow velocity may be different in
different locations, $x_1$ and $x_2$, and this influences the
correlations of the observed momenta. Thus, the scalar products
in terms of $ k$ and $ q$ become:
\be
\begin{split}
& \exp(-p_1 u_1)\, \exp(-p_2 u_2) = \\
&  \exp(-(k+q/2)\, u_1)\, \exp(-(k-q/2)\, u_2) =  \\ 
& \exp(-k u_1)\, \exp(-k u_2)\,   \exp(-q (u_1-u_2)/2) 
\end{split}
\ee
where we used the notation $u_1 = u(x_1) = u^\mu(x_1)$.
We assume that for a given detector position the normal direction
of the emission is approximately the same, so for the two
sources the term $p^\mu  \hat\sigma_\mu(x)\,$ is the same 
and it cancels in the nominator and denominator.

Thus, the expression of the correlation function, Eq. (\ref{R1}) will
be modified to
\be
\begin{split}
R(k,q) = \int\!  &  d^4 x_1 d^4 x_2\, S(x_1, k) S(x_2, k) 
    \cos[q(x_1{-}x_2)] \times  \\ 
&   \exp\left[ -\frac{q}{2} \cdot \left( \frac{u(x_1)}{T(x_1)}
                                  -\frac{u(x_2)}{T(x_2)}  
              \right)\right],
\end{split}
\label{R-def2}
\ee
and the corresponding $J(k,q)$ function will become
\be
J(k,q) =  \int d^4x\ S(x,k)\, 
\exp\left[ - \frac{q \cdot u(x)}{2T(x)} \right]\, \exp(iqx)\ ,
\label{J2} 
\ee

In Eq. (\ref{R-def2})
the term including the momentum component $q$ 
and the flow velocity $u$ becomes
unity if the source has a uniform distribution of
$u(x)/T(x)$, and in this case we may be able to use
the so called {\it smoothness approximation}:
$ S(x,k{+}q/2) S(y,k{-}q/2) \approx  S(x,k) S(y,k)$,
and the correlation function,  expression (\ref{C-def}), takes the form
\be
C( k,  q) =
1 + \frac{\left| \int d^4x\ e^{iqx} S(x,k)\right|^2}
{\left| \int d^4 x\,  S(x, k) \right|^2}\ ,
\label{C-def0}
\ee

%%%%%%%%%%%%%%%%%%%%%%%%%%%%%%%%%%%
\subsection{One Fluid Cell as Source}
%%%%%%%%%%%%%%%%%%%%%%%%%%%%%%%%%%%
\label{SSofcs}

We now assume a source function, which is reduced to one Freeze Out (FO)
time moment. Thus the integration over the 4-volume of an emission
layer is reduced to the 3-volume of a FO hypersurface. For simplicity,
we assume FO along the timelike cordinate, $t$, where we assume a
local J\"uttner distribution. Thus, we
have the source function as 
\begin{equation}
S({x}, k) =
G(x)\,H(t) 
\exp\left(-\frac{k_\mu u^\mu (x)}{T(x)} \right)
k^\mu\, \hat\sigma_\mu\ ,
\end{equation}
where $k^\mu \hat\sigma_\mu $ is an invariant scalar, and for
a single cell we use a simple quadratic parametrization for $n(x)$ as:
\begin{equation}
G(x) = \gamma n(x) = 
\gamma n_s  \exp\left( - \frac{x^2 + y^2 + z^2}{2 R^2} \right) .
\end{equation}
Here $n_s$ is the average density of the Gaussian source (or fluid cell)
of mean radius $R$.

\bigskip
%%%%%%%%%%%%%%%%%%%%%%%%%%%%%%%
{\bf Single source at rest:}
The invariant scalar $k^\mu u_\mu$ can be calculated in the frame
where the cell is at rest. 
We have then
\begin{equation}
u^\mu = (1,0,0,0) \Rightarrow 
-\frac{k_\mu u^\mu}{T} =
-\frac{k^0}{T} = 
-\frac{E_k}{T} \ .
\end{equation}
In this simplest case we also assume that the FO direction is
$\hat\sigma^\mu = (1, 0, 0, 0)$, so the $\tau$-coordinate coincides with
the $t$-coordinate, and it is orthogonal to the $x, y, x-$ coordinates.
Then we can make use of the following integral:
\begin{equation}
\int_{-\infty}^{+\infty} e^{-a x^2} d^3 x = 
\left(\frac{\sqrt \pi}{\sqrt a}\right)^3 \ .
\end{equation}
%
%\begin{equation}
%\int_{-\infty}^{+\infty} e^{-a x^2} e^{-2\pi i k x} d^3 x = 
%\left(\frac{\sqrt \pi}{\sqrt a}\right)^3 e^{-3 \pi ^2 k^2 / a} \ .
%\end{equation}
%
We can perform the integral along the $t$ direction of
$H(t)$, which gives unity and then
the single particle distribution is
\be
\begin{split}
& \int\! d^4x\ S(x, k) = \frac{n_s}{C_n} \ (k^\mu \hat\sigma_\mu) \ 
        \exp\left(-\frac{E_k}{T_s}\right) \times \\
&\int_{-\infty}^{+\infty} \!\!\!\!\!\!
                        H(t)  dt 
  \int_{-\infty}^{+\infty} \!\!\!\!
                        e^{-\frac{x^2}{2R^2} } dx 
  \int_{-\infty}^{+\infty} \!\!\!\!
                        e^{-\frac{y^2}{2R^2} } dy 
   \int_{-\infty}^{+\infty} \!\!\!\! 
                        e^{-\frac{z^2}{2R^2} } dz \ = \\
& n_s \ (k^\mu \hat\sigma_\mu) \ \exp\left(-\frac{E_k}{T_s}\right)
    \frac{\left(2 \pi R^2 \right)^{3/2} }{C_n} \ ,
\end{split}
\label{InS}
\ee
where $T_s$ is the temperature of the source, and $E_k=k^0$
in the rest frame of the fluid cell.
Due to the normalization of $H(t)$ the integral over the time $t$
is unity. The contribution to the nominator from Eq. (\ref{J2}) is
\be
\begin{split}
& 
J(k,q) = \int  d^4x\,  e^{i  q \cdot  x} e^{-q^0/(2T_s)} S({x}, k) \  = \  
\frac{n_s\,(k^\mu \hat\sigma_\mu)}{C_n} \ \times \\
&
\exp\left[{-}\frac{E_k {+} q^0/2}{T_s}\right]
\int_{-\infty}^{+\infty}\!\!\!\!\!\!\! H(t) e^{iq^0 t} dt
\int_{-\infty}^{+\infty}\!\!\!\!\!\! e^{{-\frac{x^2}{2R^2}}} e^{-iq_x x} dx                                                    \ \   \times   \\
& \ \ \ 
\int_{-\infty}^{+\infty}\!\!\!\!\!  e^{{-\frac{y^2}{2R^2}} }
                                         e^{-i q_y y} dy 
\int_{-\infty}^{+\infty}\!\!\!\!\!  e^{{-\frac{z^2}{2R^2}} }  
                                         e^{-i q_z z} dz  = \\
&
\frac{n_s (k^\mu \hat\sigma_\mu)}{C_n} \left(2 \pi R^2 \right)^{3/2} 
\exp\left[{-}\frac{E_k}{T_s}\right] \exp\left[{-}\frac{q^0}{2T_s}\right] 
\times   \\
& \ \ \ 
\exp\left[-\frac{R^2}{2} q^2\right] 
\exp\left[-\frac{\Theta^2}{2} (\hat\sigma^\mu q_\mu)^2\right] 
,
\end{split}
\ee
where we used 
$\int_{-\infty}^\infty \exp(-p^2x^2 {\pm} qx) dx = 
(\sqrt{\pi} / p) \times \exp(q^2/(4p^2)) $ \cite{GR} 3.323/2.
In the time integral the present choice of $\hat\sigma^\mu$ would
give $(q^0)^2$, but we wanted to indicate that other choices are 
also possible and they would yield $(\hat\sigma^\mu q_\mu)^2$.
In the $J(k,q) J(k,-q)$ product the terms $\exp[\pm q^0 /(2T_s)]$
cancel each other.
Inserting these equations into (\ref{C-def}) we get
\begin{equation}
C( k,  q) = 1 +
\exp\left(-\Theta^2 (\hat\sigma^\mu q_\mu)^2 - R^2 q^2\right) \ .
\label{Csst}
\end{equation}
If we have a source at a point in the FO layer, which is at a longer 
distance from the external side of the FO layer than $\Theta$, then
the contribution of the time integral from this point is reduced.
In a few source model it is more transparent to describe this reduction 
by assigning a smaller weight factor to the contribution of the deeper 
lying source. See in section \ref{ASm}.

\begin{figure}[ht] %%%%%%%%%%
\begin{center}
      \includegraphics[width=7.6cm]{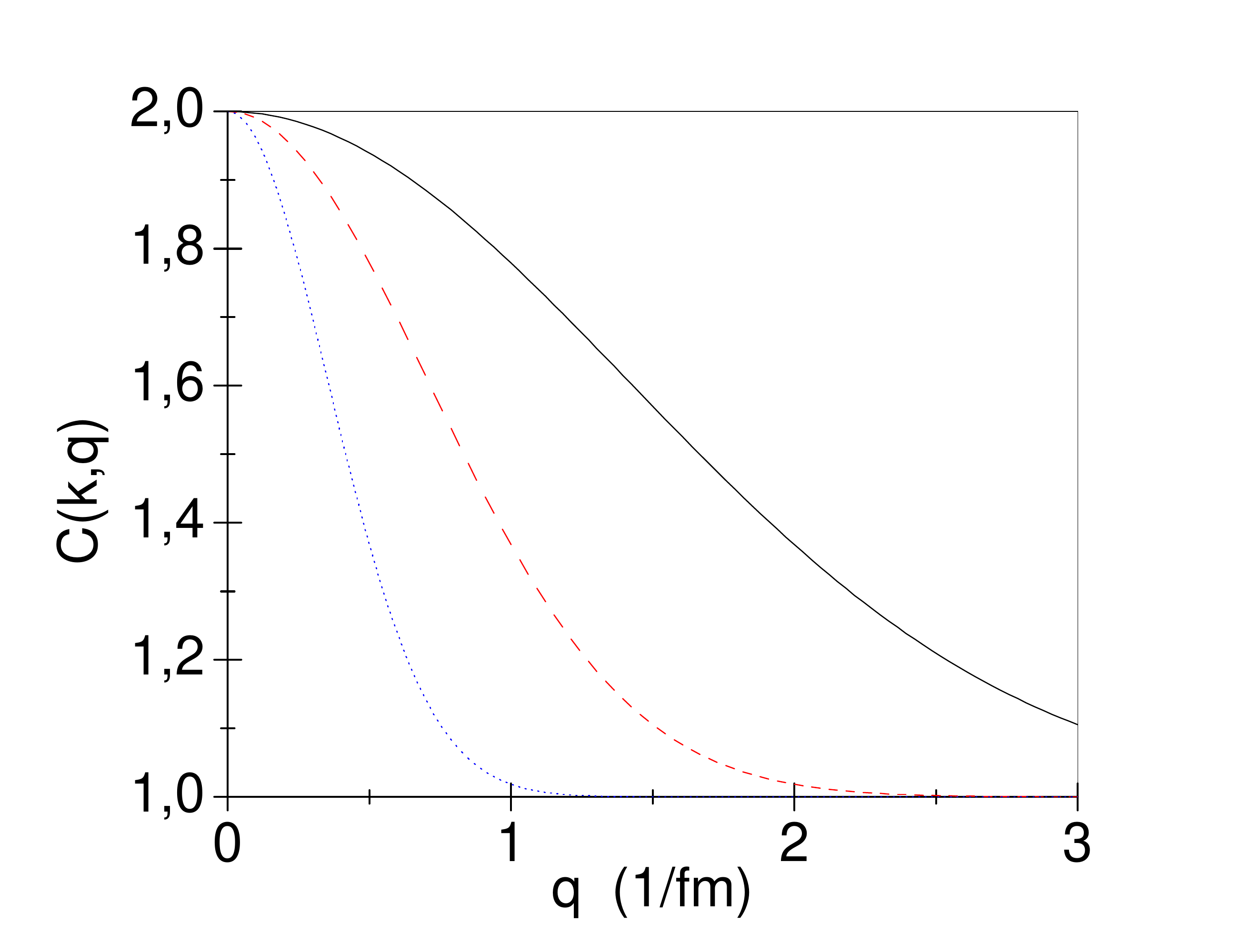}
\end{center}
\vskip -8mm
\caption{ (color online) 
The correlation function, $C(k,q)$, for a single, static, 
spherically symmetric,
Gaussian source with different radii, $R= 4, 1$ and $0.25$ fm, 
({\it blue dotted, red dashed,} and {\it full black lines} 
respectively), as described by Eq. (\ref{Csss}). }
\label{F-1}
\end{figure}

If we tend to an infinitely narrow FO layer, 
$\Theta \rightarrow 0$, i.e. to a FO hypersurface, then
\begin{equation}
C( k,  q) = 1 +
\exp\left(-R^2 q^2\right) \ .
\label{Csss}
\end{equation}
The $ k$ dependence drops out from the correlation function, 
$C( k,  q)$ as the $ k$ dependent parts are separable. 
See Fig. \ref{F-1}.
The size of the fluid cells in a high resolution 3+1D fluid dynamical
calculation is $(0.3$fm$)^3$. With this resolution the
{\it numerical viscosity} of the fluid dynamical calculation 
\cite{Horvat} is the same as the estimated minimal viscosity
of the QGP \cite{Kovtun} which occurs at the critical point
of the phase transition \cite{CsKM}. As Fig. \ref{F-1} shows
the correlation for such a cell size yields to an extended 
distribution in the relative momentum $q$.

For the study of the rotation of the system the thickness of the
FO layer is of secondary importance, especially if we discuss only
a few fluid sources. In this case the role of the depth of a source
point within the layer is given by its reduced contribution to the
particle emission. This can be represented much simpler with assigning
emission weights to the small number of sources. Thus, in the following
discussion, we do not go into the details of the time structure of the
emission.

\bigskip 
%%%%%%%%%%%%%%%%%%%%%%%%%%
{\bf Single moving source:}
Let us take a single source which moves
in the x-direction with a velocity $v_x$. Then we have,
$  u_s^\mu = \gamma_s (1,v_x,0,0) $, and the scalar product
$k\cdot u_s/T_s = k_\mu u_s^\mu / T_s $ provides an additional
contribution to the correlation function. However, in the case
of a single fluid cell or a single source the
velocity and the temperature do not change within
the cell, so the modifying term in
eq. (\ref{R-def2}) becomes unity.
We use $k_\mu u^\mu = \gamma (E_k - k_x v_x)$, and the source function
becomes
\begin{equation}
S({x}, k) =
\frac{n(x) \, (k^\mu\,\hat\sigma_\mu)}{C_n} \, 
\exp\left[-\frac{k \cdot u_s}{T_s}\right] \ ,
\end{equation}
where $[k\cdot u_s/T_s ] = [\gamma_s (E_k - k_x v_x)/T_s]$.

Within the source (or fluid element) the velocity $u_s$ and temperature
$T_s$ are assumed to be the same. The source or fluid element
may have a density profile, but this profile should be the same
for all cells (although the average density, $n_s$ is not the same for
all cells. The spatial integrals can be performed in the
rest frame of the cell, giving the same integral result as above 
(\ref{InS}), because the moving cell-size shrinks, but the apparent
density increases, so that the total number of particles in a 
cell remains the same as it is an invariant scalar. 

\begin{figure}[ht] %%%%%%%%%%
\begin{center}
      \includegraphics[width=6cm]{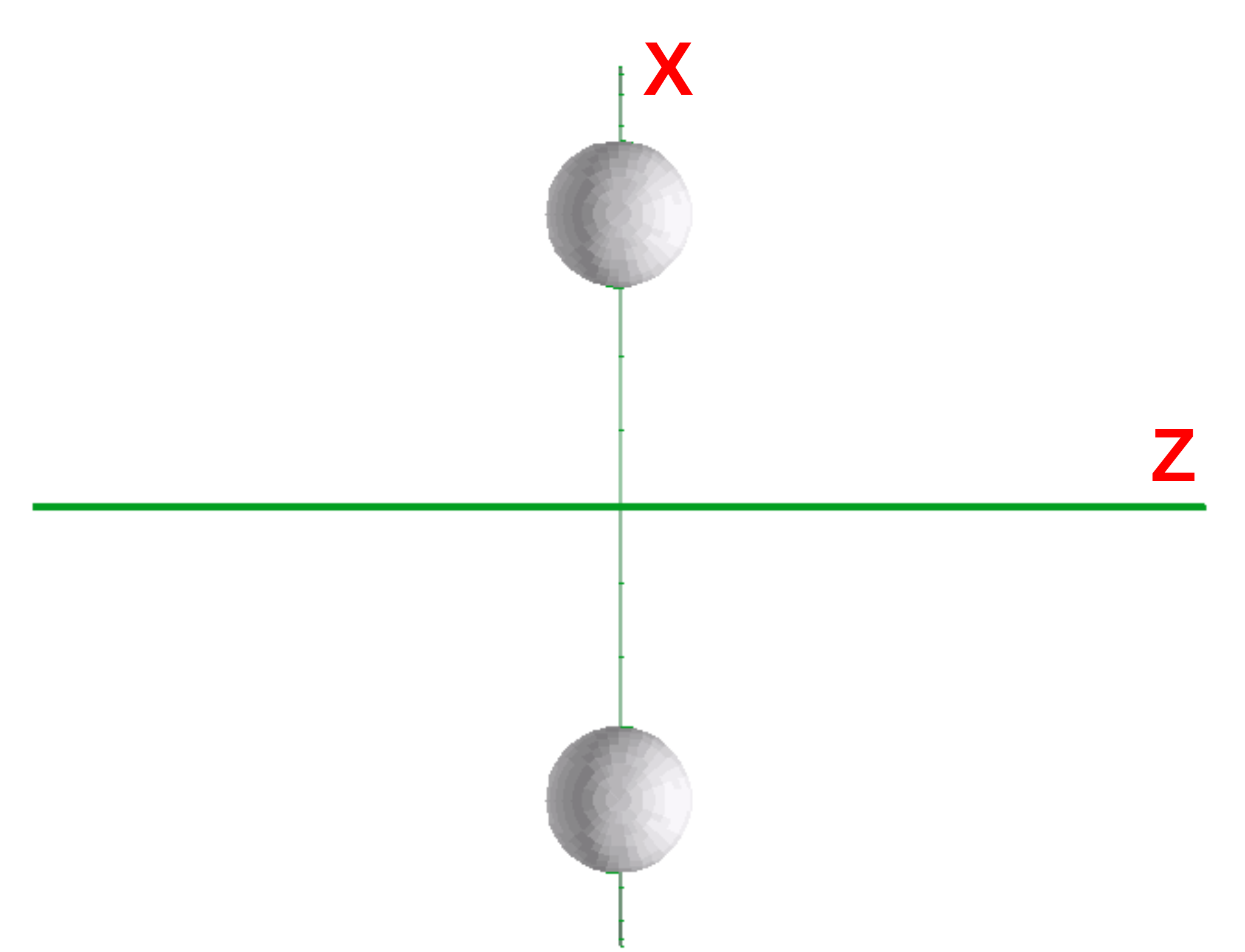}
\end{center}
\vskip -4mm
\caption{ (color online)
Two steady sources in the reaction ($[x-z]$) plane with a distance
between them of $2d$ in the $x-$direction.}
\label{F-2}
\end{figure}

Then the integral of the single particle contribution is
\begin{equation}
\begin{split}
\int\! d^3 x\,  S({x}, k) \, & =   \\
 n_s (k^\mu \hat\sigma_\mu) &
 \exp\left[-\frac{k \cdot u_s}{T_s}\right] \,
    \frac{\left(2 \pi R^2 \right)^{3/2} }{C_n} .
\end{split}
\end{equation}

%%%%%%%

Then the two particle distribution:
\begin{equation}
\begin{split}
     J(k,q) =  \int\! d^3 x\, e^{i  q \cdot  x} S({x}, k) 
  &\exp\left[ - \frac{q \cdot u_s}{2T_s} \right]\, = \\
     n_s (k^\mu \hat\sigma_\mu) \exp\left[-\frac{k \cdot u_s}{T_s}\right] 
  &\exp\left[ - \frac{q \cdot u_s}{2T_s} \right] \times \\
    \frac{\left(2 \pi R^2 \right)^{3/2} }{C_n} 
  &\exp\left(-\frac{R^2}{2} q^2\right) \ .
\end{split}
\end{equation}
When calculating $R(k,q)$, in the $J(k,q) J(k,-q)$ product 
the terms $\exp[\pm q \cdot u_s /(2T_s)]$ cancel each other.
In the formulae the $\hbar = 1$ convention is used and $k$ and $q$ are
considered as the wavenumber vectors.

We then insert these equations into equation (\ref{C-def}) and we get for one 
moving Gaussian source
\begin{equation}
C( k,  q) =
1 + \exp\left(-R^2 q^2\right) \ .
\end{equation}
Again, this result does not depend on $ k$, just as the 
previous single source at rest, Eq. (\ref{Csss}).

\bigskip

%%%%%%%%%%%%%%%%%%%%%%%%%%%%%%%%%%%
\section{Symmetric Few Source Models}
%%%%%%%%%%%%%%%%%%%%%%%%%%%%%%%%%%%
\label{FSM}

%%%%%%%%%%%%%%%%%%%%%%%%%%%%%%%%%%%
\subsection{Two Steady Fluid Cells}
%%%%%%%%%%%%%%%%%%%%%%%%%%%%%%%%%%%

For emission from two steady sources,
two particle correlations were studied in 
ref. \cite{Cso-5}. Here we use the present method. We assume that 
the two source system is symmetric both their positions are placed
symmetrically and 
also their FO normal vectors, $\hat\sigma^\mu$, are the same.
If the normal were $\hat\sigma^\mu =(1,0,0,0)$, then the
invariant scalar $k^\mu \hat\sigma_\mu$ would be $k^0 = E_k$,
although we do not need this additional requirement to illustrate
the correlation function, which would arise from an idealized symmetric 
system. 

We also assume that the time distributions, $H(\tau)$ for the
two sources are identical, so these can be integrated simultaneously 
and yield unity.  If we have two
sources then the source function is
\be
\begin{split}
&  S(x, k) = \sum_s S_s(x,k) = \\
&  (k^\mu \hat\sigma_\mu) \sum_s \frac{n_s(x) \,}{C_{ns}} \, 
\exp\left[-\frac{k \cdot u_s}{T_s}\right] \ ,
\end{split}
\ee
while the $J$ function in the J\"uttner approximation is
\be
\begin{split}
& J(k,q) = \\
& \sum_s 
\exp\left[ - \frac{q \cdot u_s}{2T_s} \right]\, \exp(iqx_s)
\int_S d^4x\ S_s(x,k)\, 
 \exp(iqx)\ ,
\end{split}
\label{J2s} 
\ee
where $x_s$ is the position of the center of the source, 
and the spatial integrals run separately for each of the 
identical sources, i.e. we assume fluid cells with identical density
profiles, but with different densities, $n_s$ and temperatures, $T_s$.

In case of steady sources $u^\mu_s = (1,0,0,0)$, and the spatial integral 
for one source is the same as for a single source. Thus,
\be
\begin{split}
   \int\! d^3x\ &  S(x, k) = \sum_s \int_S d^3x\ S_s(x,k) = \\ 
&  \left(2 \pi R^2 \right)^{3/2} \, 
   (k^\mu \hat\sigma_\mu) \sum_s  \frac{n_s }{C_{ns}} 
   \exp\left(-\frac{E_k}{T_s}\right)
\end{split}
\label{S2sm}
\ee 
and 
\be
\begin{split}
& J(k,q) = \\
&  \sum_s \exp\left[-\frac{q^0}{2 T_s} \right]\, \exp(iqx_s)
   \int_S d^3x\ S_s(x,k)\, \exp(iqx)  = \\
&  \left(2 \pi R^2 \right)^{3/2} \ (k^\mu \hat\sigma_\mu) \ 
   \exp\left(-\frac{R^2}{2} q^2\right) \  \sum_s \frac{n_s}{C_{ns}}\ \times \\
&  \exp\left(-\frac{E_k}{T_s}\right) 
  \exp\left[-\frac{q^0}{2 T_s} \right]\, 
  \exp(iq^0 x^0_s) \exp(-i \vec q \vec x_s) \ .
\end{split}
\label{J2sM} 
\ee
In the $J(k,q) J(k,-q)$ product the terms $\exp[\pm q^0 /(2T_s)]$
cancel each other. Both $J(k,q)$ and $J(k,-q)$ include a sum
$[\exp(i\vec q \vec x_s) + \exp(-i\vec q \vec x_s)]$, 
and their product leads to a
factor $2[ 1 + \cos(2 \vec q \vec x_s)]$. 
Here we assumed that the time-like extent of the emission layer
is negligible compared to the space-like size.

Consequently, if the two sources have the same parameters, just
different locations, $x_1 = -x_2$ 
(see Fig. \ref{F-2}) then
\be
C(k,q) = 1+ \frac{1}{2} \exp(-R^2 q^2) [1+ \cos(2 \vec q \vec x_s)]
\label{C2}
\ee

\begin{figure}[ht] %%%%%%%%%%
\begin{center}
      \includegraphics[width=7.6cm]{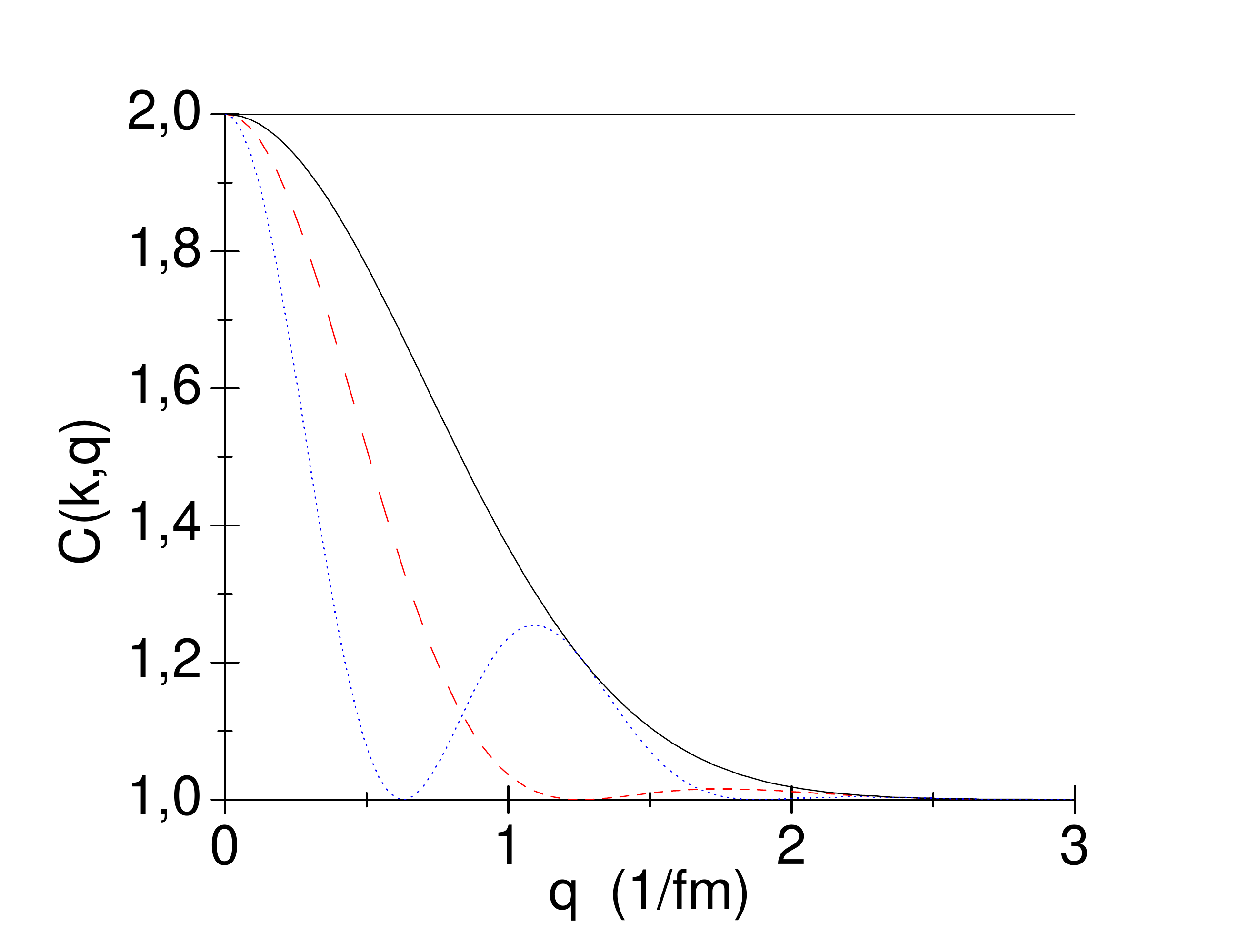}
\end{center}
%\vskip -8mm
\caption{ (color online)
The correlation function, $C(k,q)$, for two spherical, 
Gaussian sources at rest, shown in the direction of the 
displacement (here $q_x$) with different 
distances, $d = 2.5, 1.25$ and $0$ fm, 
({\it dotted blue, dashed red,} and {\it full-black lines} 
respectively)
and in the orthogonal direction, $q_y$ which is identical with
the $d=0$ contribution ({\it full black line}). For finite $d$
the cosine term results in a modification in the direction of 
the line joining the sources, which becomes apparent if the
distance between the sources becomes small compared to the
size of the spherical sources.}
\label{F-3}
\end{figure}
\bigskip

This result agrees with ref. \cite{Cso-5}, section 9.1 (p. 41),
and in the limit of $\vec x_s = 0$ it returns the single source
result, Eq. (\ref{Csss}). 
See Fig. \ref{F-3}.
If the distance of the two sources is 
$2d$, i.e. $x_1=d$ and $x_2=-d$, then $2 \vec q \vec x_s = 2 q_x\, d$,
thus the modification appears in the $q_x$-direction only. In the
other directions, $q_y$ and $q_z$, the single source result
(\ref{Csss}) is returned.  

If the distance of the two sources, $2d$, is comparable or smaller than
the radius of a single source, $R$, then the two source configuration
leads to visible zero points, $C(k,q) = 0$, on the $q_x$-axis at 
$2 q_x\, d= \pm (1+2n) \pi$, where $n= 0, 1, 2, 3, ...$ \, . In Fig.
\ref{F-3} for the $d=2.5$fm case we see these zero points at 
$q_x = \pi/(2d), 3\pi/(2d)\,,\,$ ... , while at the points 
$q_x = 2\pi/(2d), 4\pi/(2d)\,,\,$ ... the distribution function,
$C(k,q_x)$ touches (becomes tangent to) the distribution function
for $d=0$ or the distribution function $C(k,q_y)$.

The appearance of the zero points  is to a large extent an artifact 
of the used very simplistic two source model. In case of other additional 
sources these zero points would disappear. Nevertheless, this feature
illustrates that the correlation function can be more complex than
a set of Gaussians of the momentum difference $q$ in different directions
or at different rapidities.

\begin{figure}[ht] %%%%%%%%%%
\begin{center}
      \includegraphics[width=6cm]{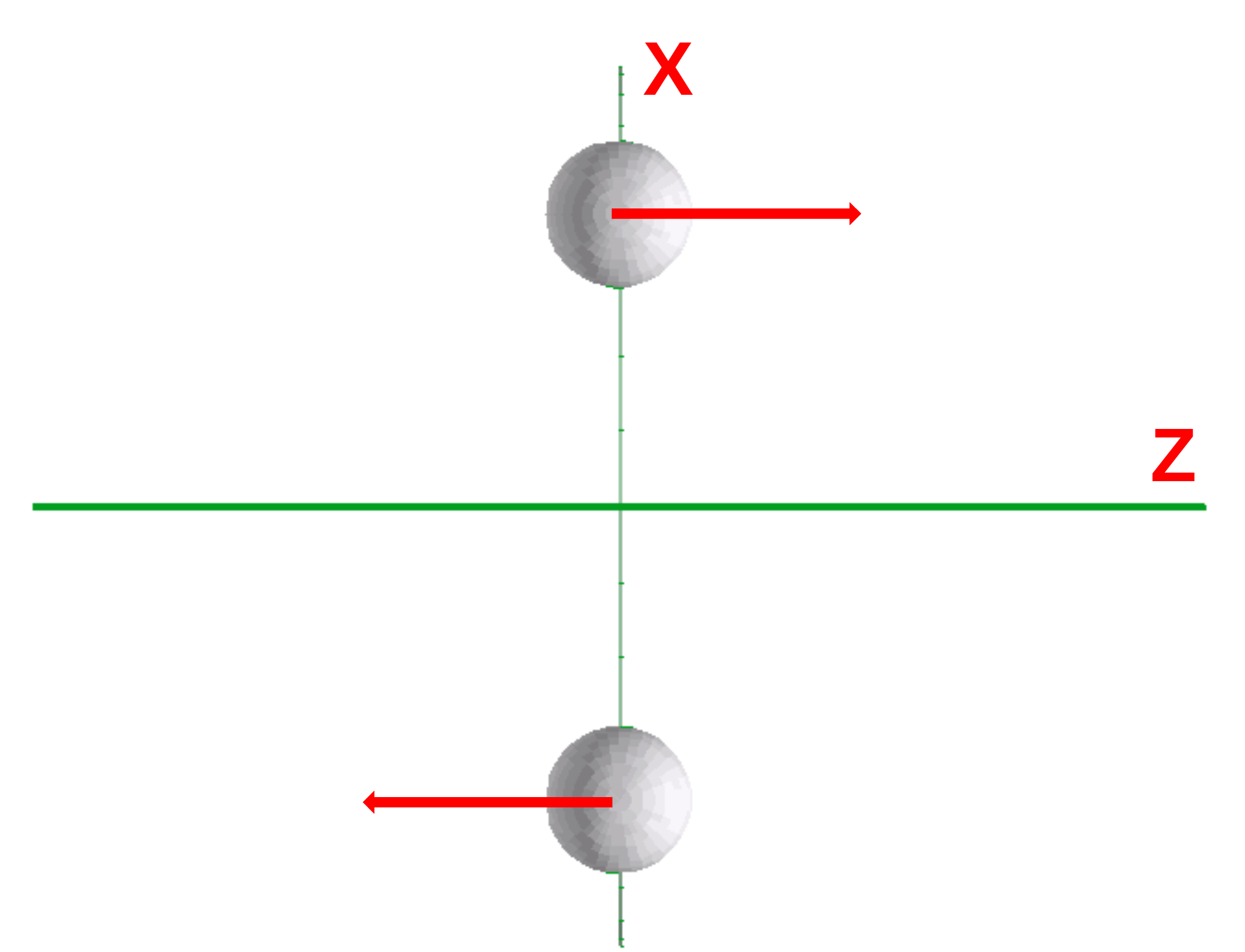}
\end{center}
%\vskip -4mm
\caption{ (color online)
Two moving sources in the reaction ($[x-z]$) plane with a distance
between them of $2d$ in the $x-$direction. The sources are moving in the
directions indicated by the (red) arrows.}
\label{F-4}
\end{figure}

%%%%%%%%%%%%%%%%%%%%%%%%%%
\subsection{Two moving sources}\label{Ltsm} 
%%%%%%%%%%%%%%%%%%%%%%%%%%

We study the system the same way as before, but now the two sources 
are moving in opposite directions, so that
$u_s = u_1$ or $u_2$ where
$u_1^\mu = (\gamma_s, \gamma_s \vec v_1)$, 
$u_2 = \bar u_s^\mu = (\gamma_s, \gamma_s (- \vec v_1))$,   and
$\vec u_s \equiv \gamma_s\, \vec v_s$,      so that
$\vec u_1 = - \vec u_2$, see Fig. \ref{F-4}. Similarly,
$x_s = x_1$ or $x_2$ where
$x_s^\mu = (t_s, \vec x_s)$, 
$\bar x_s^\mu = (t_s, - \vec x_s)$,
and $\vec x_1 = - \vec x_2$.  For now we also assume that
FO happens at a $t=$const. FO hypersurface, so 
$d\hat\sigma^\mu = (1, 0,0,0)$ and so $t_1 = t_2$.

If we have 
several sources then the source function in J\"uttner approximation is

\be
\begin{split}
& S({x}, k) = \sum_s S_s(x,k) = \\
& (k^\mu\, \hat\sigma_\mu) \ \sum_s \frac{n_s(x) \, }{C_{ns}} \, 
\exp\left[-\frac{k \cdot u_s}{T_s}\right] \ ,
\end{split}
\ee
while the $J$ function is
\be
\begin{split}
 J(k,q) = & \sum_s  
         \exp\left[- \frac{q \cdot u_s}{2T_s}\right] \times \\
& \exp(iqx_s) \int_S d^3x\ S_s(x,k)\, \exp(iqx)\ ,
\end{split}
\ee

where $x_s$ is the 4-position of the center of source $s$, 
and the spatial integrals run separately for each of the 
identical sources, i.e. we assume fluid cells with identical density
profiles, but with different densities, $n_s$,
velocities, $u_s$ and temperatures, $T_s$.

The spatial integral 
for one source is the same as for a single source. Thus,

\begin{widetext}

\be
 \int\! d^3x\   S(x, k) = \sum_s \int_S d^3x\ S_s(x,k) = 
  (k^\mu\, \hat\sigma_\mu) \ \left(2 \pi R^2 \right)^{3/2} \ 
   \  \frac{n_s}{C_{ns}} \exp\left(-\frac{k^0 \gamma_s}{T_s}\right) 
 \left[ 
\exp\left( \frac{\vec k \vec u_s}{T_s} \right)+ 
\exp\left(-\frac{\vec k \vec u_s}{T_s} \right) 
\right] \ .
\ee
This returns Eq. (\ref{S2sm}) if $u^\mu_s = (1,0,0,0)$.
The function $J(k,q)$ becomes
\be
\begin{split}
& J(k,q) = 
 \sum_s \exp\left[-\frac{q \cdot u_s}{2 T_s} \right]\, 
\exp(iqx_s)  \int_S d^4x\ S_s(x,k)\, \exp(iqx)  = \\
& 
     (k^\mu\, \hat\sigma_\mu) \ \left(2 \pi R^2 \right)^{3/2}  
\exp\left(-\frac{R^2 q^2}{2} \right) 
  \sum_s \frac{n_s}{C_{ns}} \exp\left[-\frac{k \cdot u_s}{T_s}\right] 
  \exp\left[-\frac{q \cdot u_s}{2 T_s} \right]\, \exp(iqx_s)  = \\
& 
      (k^\mu\, \hat\sigma_\mu) \ \left(2 \pi R^2 \right)^{3/2} 
\exp\left(-\frac{R^2}{2} q^2\right) \frac{n_s}{C_{ns}}
\exp\left[- \frac{k^0 \gamma_s}{T_s}\right]
\exp\left[- \frac{q^0}{2} \frac{\gamma_s}{T_s}\right]
\exp(i q^0 x_s^0) \times \\
& \left[
  \exp\left[ \frac{\vec k \vec u_s}{  T_s}  \right] 
  \exp\left[ \frac{\vec q \vec u_s}{2 T_s}  \right]
           \exp(-i \vec q \vec x_s) + 
  \exp\left[ -\frac{\vec k \vec u_s}{  T_s}  \right] 
  \exp\left[ -\frac{\vec q \vec u_s}{2 T_s}  \right]  
           \exp(i \vec q \vec x_s) \right],
\end{split}
\label{J2s3M} 
\ee
where the factor $\exp(i q^0 x_s^0)$ can be dropped if the FO time 
distribution is simultaneous for the two sources,
because then $x_s^0 = 0$.
This returns Eq. (\ref{J2sM}) if $u^\mu_s = (1,0,0,0)$. 

Now we can divide the two particle correlation with the
square of the single particle distribution

\be
\begin{split}
\frac{ Re\, [ J(k,q)\, J(k,-q) ] }{\left| \int d^4 x\,  S(x, k) \right|^2} 
&
\ \ = \ \ 
  \exp( -R^2 q^2) 
\frac{ Re \left[
e^{\frac{2 \vec k \vec u_s}{T_s}} + e^{-\frac{2 \vec k \vec u_s}{T_s}} +
e^{ \frac{\vec q \vec u_s}{T_s}} e^{ i 2 \vec q \vec x_s} +
e^{-\frac{\vec q \vec u_s}{T_s}} e^{-i 2 \vec q \vec x_s}
 \right] }{
\left(e^{ \frac{\vec k \vec u_s}{T_s}} +
      e^{-\frac{\vec k \vec u_s}{T_s}} \right)^2 } \\
=   \exp( - R^2 q^2) &
\frac{
\ Re\, \left[
2 \cosh\left( \frac{2 \vec k \vec u_s}{T_s} \right) +
 e^{ \frac{\vec q \vec u_s}{T_s} } 
\left(\cos( 2\vec q \vec x_s) + i \sin( 2\vec q \vec x_s) \right) +
 e^{ -\frac{\vec q \vec u_s}{T_s} } 
\left(\cos(-2\vec q \vec x_s) + i \sin(-2\vec q \vec x_s) \right) 
\right]
}{
2 \left[ \cosh\left( \frac{2 \vec k \vec u_s}{T_s} \right) +1 \right] } \\
=   \exp( -R^2 q^2) &
\frac{
 \cosh\left( \frac{2\vec k \vec u_s}{T_s} \right) +
 \cosh\left( \frac{ \vec q \vec u_s}{T_s} \right) 
  \cos( 2\vec q \vec x_s)  
}{  \cosh\left( \frac{2 \vec k \vec u_s}{T_s} \right) +1  } \\
\end{split}
\label{C2msd}
\ee

\end{widetext}

Consequently, if the two sources have the same parameters, just
opposite locations with respect to the center, and opposite velocities,  
then the correlation function is

\be
\begin{split}
 C(k,q) & =  1 +   \exp(-R^2 q^2) \ \times \\ 
& 
\frac{
 \cosh\left( \frac{2\vec k \vec u_s}{T_s} \right) +
 \cosh\left( \frac{ \vec q \vec u_s}{T_s} \right) 
  \cos( 2\vec q \vec x_s)  
}{  \cosh\left( \frac{2 \vec k \vec u_s}{T_s} \right) +1  } \ .
\end{split}
\label{C2ms}
\ee
This returns Eq. (\ref{C2}) if $u^\mu_s = (1,0,0,0)$, and $C(k,q) = 2$
if $q = 0$.
\bigskip

If we have two sources placed at $ x = \pm d_x$, 
and with the velocity in the $\pm z$-direction, $\pm v_z$, 
then the correlation function is for the different directions 
becomes:

%%%

\begin{equation}
\begin{split}
C(k_x,q_x) & = 1 + \frac{1}{2} \exp(-R^2 q^2_x)
\left[1+\cos\left(2 q_x d_x \right) \right]\ , \\
C(k_x,q_y) & = 1 + \exp(-R^2 q^2_y)\ , \\
C(k_x,q_z) & = 1 + \frac{\exp(-R^2 q^2_z)}{2}
\left[ 1+ \cosh\left( \frac{\gamma q_z  v_z}{T_s} \right) \right]\ .
\end{split}
\label{C2kx}
\end{equation}

%%%

\begin{equation}
\begin{split}
C(k_y,q_x) & = 1 +  \frac{1}{2} \exp(-R^2 q^2_x)
\left[ 1+ \cos\left( 2 q_x  d_x \right)  \right]\ , \\
C(k_y,q_y) & = 1 + \exp(-R^2 q^2_y)\ , \\
C(k_y,q_z) & = 1 + \frac{\exp(-R^2 q^2_z) }{2}
\left[ 1+ \cosh\left( \frac{\gamma q_z  v_z}{T_s} \right) \right]\ .
\end{split}
\end{equation}

\begin{figure}[ht] %%%%%%%%%%
\begin{center}
      \includegraphics[width=7.6cm]{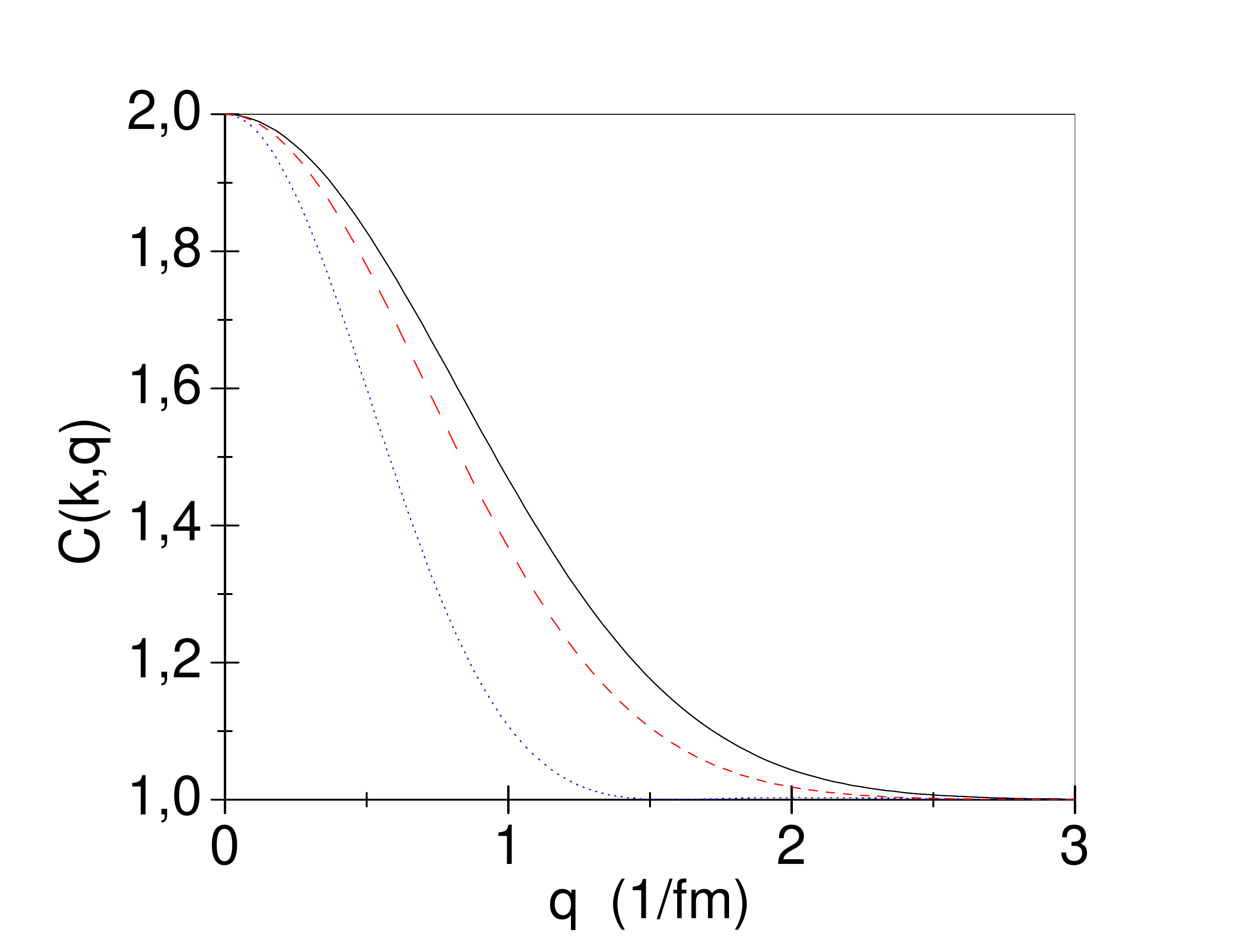}
\end{center}
\vskip -8mm
\caption{ (color online)
The correlation functions, $C(k,q)$, for two moving sources where the 
displacement of the sources is in the $x$-direction, and the
center-of-mass momentum, $\vec k$, of emitted particles is in the 
$x$ and $y$ direction. The {\it solid black line} is for the momentum 
difference, $q_z$, the {\it dashed red line} is 
for $q_y$ and {\it dotted blue line} is for $q_x$. The radius of the
sources is $R=1$ fm, (same as in Fig. \ref{F-3}), the displacement is
$d=1$fm, and the source velocity is, $\gamma v_z /T_s = 1.0$ fm.
This can be satisfied e.g. by $u_s = 0.6$\,c and $T=0.12$ GeV.  }
\label{F-5}
\end{figure}

%%%

\begin{equation}
\begin{split}
C(k_z,q_x)  =\ & 1 + \exp(-R^2 q^2_x) \ \times \\ 
& 
\frac{
 \cosh\left( \frac{2 \gamma k_z  v_z}{T_s} \right) +
 \cos\left( 2 q_x  d_x \right)   
}{  \cosh\left( \frac{2 \gamma k_z  v_z}{T_s} \right) +1  } \ , \\
C(k_z,q_y)  =\ & 1 + \exp(-R^2 q^2_y)\ , \\
C(k_z,q_z)  =\ & 1 + \exp(-R^2 q^2_z) \ \times \\ 
& 
\frac{
 \cosh\left( \frac{2 \gamma k_z  v_z}{T_s} \right) +
 \cosh\left( \frac{\gamma q_z  v_z}{T_s} \right)   
}{  \cosh\left( \frac{2 \gamma k_z  v_z}{T_s} \right) +1  } \ .
\end{split}
\end{equation}

%%%

Therefore only the correlation functions in the $k_z, q_x$-direction 
and in the $q_z$-directions are 
affected by the $z$-directed velocity of the source. 
In this direction, $k_z$, unfortunately it is difficult to detect
the two particle correlations. 
For the $k_x$ and $k_y$-directions the $q_x$-distribution
is affected by the displacement of the two sources by $\pm d_x$. 
The $q_y$-distribution is not effected by either the displacement 
or the source velocities.

\begin{figure}[ht] %%%%%%%%%%
\begin{center}
      \includegraphics[width=7.6cm]{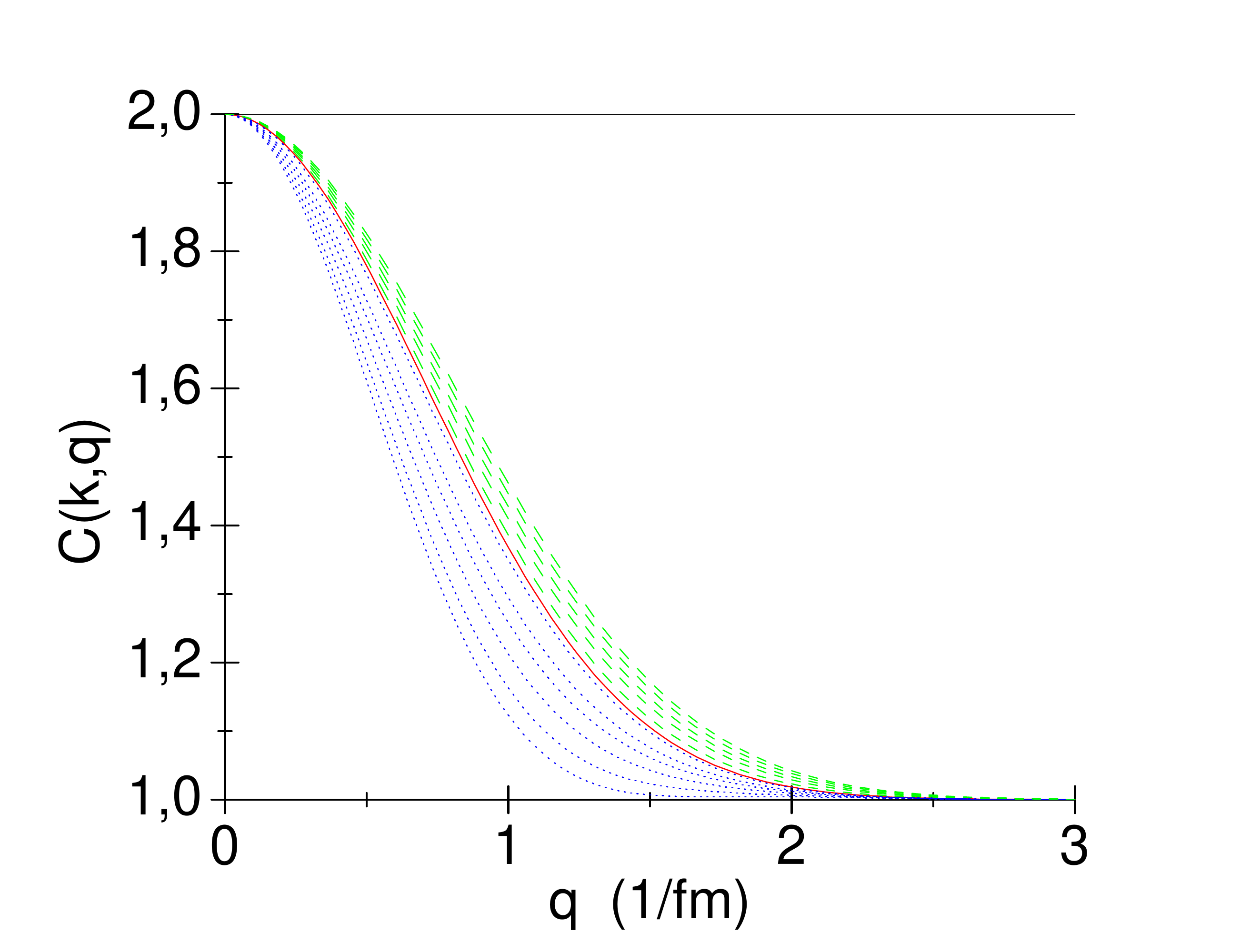}
\end{center}
\vskip -8mm
\caption{ (color online)
The correlation functions, $C(k,q)$ for two moving sources where the 
displacement of the sources is in the $z$-direction, and the 
center-of-mass momentum, $\vec k$, of emitted particles is in the 
 $x$-direction.
The {\it dashed green lines} are 
for the relative momentum, $q_z$, the {\it solid red line} is for $q_y$
and {\it dotted blue lines} are for $q_x$. 
For large values of the center-of-mass momentum $k_x$ 
the correlation functions $C(k_x,q_x)$ and $C(k_x,q_z)$ 
will approach the correlation
function $C(k_x,q_y)$ \,({\it red line}). 
For $q_x$ ({\it blue lines}) the displacements are  
$d_x = 1,0$ fm, 
and for $q_z$ ({\it green lines}) the velocity is chosen such that
$\gamma v_z /T_s = 1.0$ fm. The values of $k_z$ are for the 
{\it blue lines}: 0.25, 0.5, 0.75, 1.0, 1.25 and 2.0 fm$^{-1}$ and for the 
{\it green lines}: 0.25, 0.5, 0.75, 1.0 and 1.5 fm$^{-1}$.}
\label{F-6}
\end{figure}

The correlation function for different source locations and velocities 
are similar. The cosine term appears in the same direction as the axis at 
which the sources are located and the hyperbolic cosine in the 
direction of the velocity. See Figs. \ref{F-5} and \ref{F-6}. 
The zero points discussed for the
two static sources at Eq. (\ref{C2}), appear in the distributions
$C(k_x,q_x)$ and $C(k_y,q_x)$. These distributions do depend on the
magnitude of the flow velocity, $v_z$, but not on its direction!
This arises from the fact that the detectors are {\bf assumed to be}
reached from both sides of the system with opposite velocities
with equal probability.

Unfortunately the dominant direction of flow (see Fig. \ref{F-8}) 
is the beam direction ($z-$direction), where we have no possibility
to place high acceptance detectors. At the same time the strongest
effect of the flow appears in this direction.

\begin{figure}[ht] %%%%%%%%%%
\begin{center}
      \includegraphics[width=6cm]{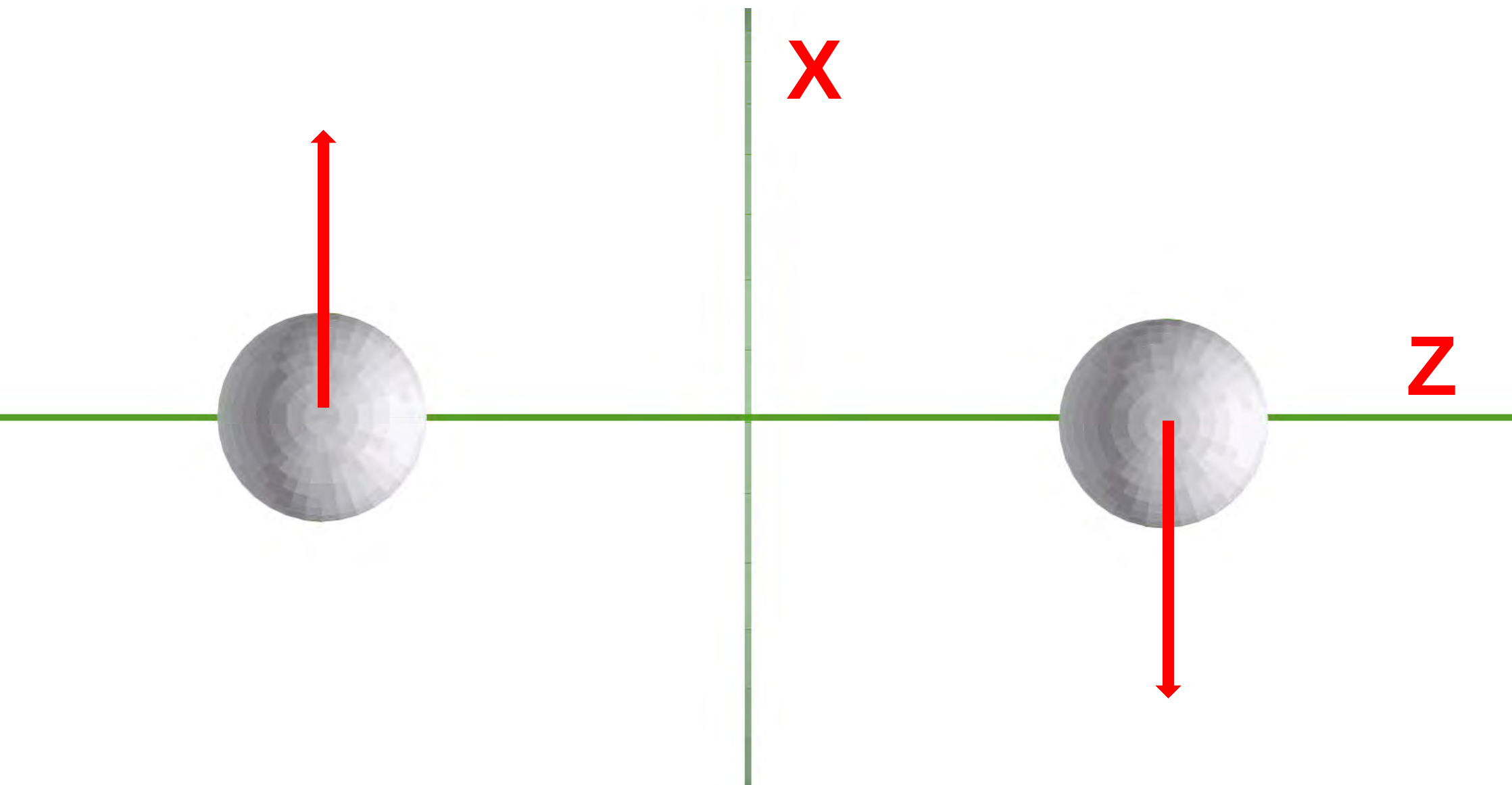}
\end{center}
%\vskip -4mm
\caption{ (color online)
Two moving sources in the reaction ($[x-z]$) plane with a distance
between them of $2d$ in the $z-$direction. The sources are moving in the
directions indicated by the (red) arrows.}
\label{F-7}
\end{figure}

\begin{figure}[ht] %%%%%%%%%%
\begin{center}
      \includegraphics[width=7.6cm]{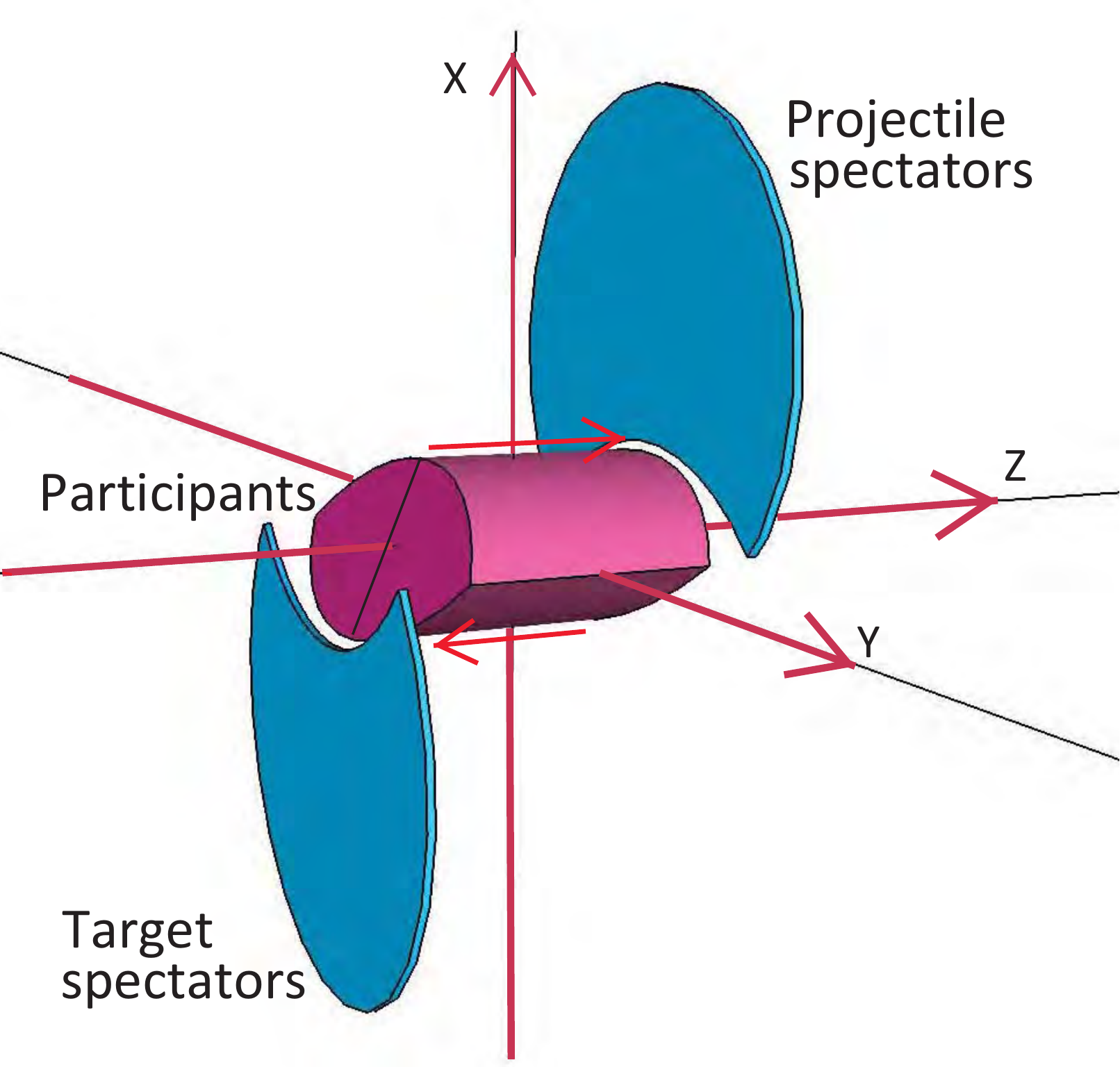}
\end{center}
%\vskip -4mm
\caption{
(Color online)
Typical orientation of the spatial axes in case of an 
ultra-relativistic heavy ion reaction shortly after the
impact. In the configuration space the projectile and
target appear to be flat due to the Lorentz contraction.}
\label{F-8}
\end{figure}

The rotation in the reaction plane can also be characterized 
with {\bf another configuration} of the two moving 
sources, when the displacement
is in the $z$-direction while the flow velocities are pointing into
the $x$-direction, so that the source at $x_1 = d_z$ has a 
negative velocity, $-v_x$ while the source at $x_2 = - d_z$ has a 
positive velocity, $v_x$. 
(See Fig. \ref{F-7}.)
The detailed description of the correlation 
functions from this configuration can be obtained in a straightforward
way similarly to the previous case, see Eq. (\ref{Oth-conf}).  
In this case the flow has
the most dominant effect in the $k_x$-direction, which is accessible
for detection. The $x$-directed flow, however, is more sensitively
dependent on secondary effects, like the Kelvin-Helmholtz Instability
\cite{hydro2}.

In this configuration of the sources the magnitude of the flow velocity
makes visible change in $C(k,q)$, in the $(k_x, q_x)$-direction also, 
which is
detectable by the usual detector configurations. Still the direction 
of the rotation does not appear in the observables with the approach 
presented here. 

This actually arises from the simplifying assumption,
that the freeze out is happening instantly at a timelike hypersurface
with $\hat\sigma^\mu = (1,0,0,0)$, where particles from all sides
of the system can reach each detector with the same probability.
We will return to this problem after having discussed the more complex
source configurations.  

%%%

\begin{equation}
\begin{split}
C(k_x,q_x)  =\ & 1 + \exp(-R^2 q^2_x) \ \times \\ 
& 
\frac{
 \cosh\left( \frac{2 \gamma k_x  v_x}{T_s} \right) +
 \cosh\left( \frac{\gamma q_x  v_x}{T_s} \right)   
}{  \cosh\left( \frac{2 \gamma k_x  v_x}{T_s} \right) +1  } \ , \\
C(k_x,q_y)  =\ & 1 + \exp(-R^2 q^2_y)\ , \\
C(k_x,q_z)  =\ & 1 + \exp(-R^2 q^2_z) \ \times \\ 
& 
\frac{
 \cosh\left( \frac{2 \gamma k_x  v_x}{T_s} \right) +
 \cos\left( 2 q_z  d_z \right)   
}{  \cosh\left( \frac{2 \gamma k_x  v_x}{T_s} \right) +1  } \ .
\end{split}
\label{Oth-conf}
\end{equation}

%%%

For these two-particle correlation measurements it is necessary to
identify independently, event by event the global collective 
reaction plane azimuth, $\Psi_{RP}$, experimentally and the
corresponding  event by event center of mass of the system
(e.g. with the method \cite{Eyyubova}). Knowing these we can
identify the $k_x$-direction (and the $k_y$-direction also.

In this section we derived a relatively simple formula
for two sources with opposite positions and opposite velocities.
These kind of systems were analysed earlier for radially expanding
systems.

Recently due to the angular momentum in peripheral heavy ion 
collisions strong rotation \cite{hydro1} 
and turbulence (Kelvin-Helmholtz Instability) \cite{hydro2}
were predicted in fluid dynamical models arising from the
symmetries, shear and vorticity of the initial state.

In the simple two source example shown in the previous section
the two sources may describe a rotation if the sources are
at a distance from the center in the x-direction,
$\vec x_1 = (+d, 0, 0)$ and $\vec x_2 = (-d, 0, 0)$,
while these have opposite velocities pointing into the
z-direction,
$\vec u_1 = \gamma (1, 0, 0, v_z)$ and 
$\vec u_2 = \gamma (1, 0, 0,-v_z)$.

\begin{figure}[ht] %%%%%%%%%%
\begin{center}
      \includegraphics[width=7cm]{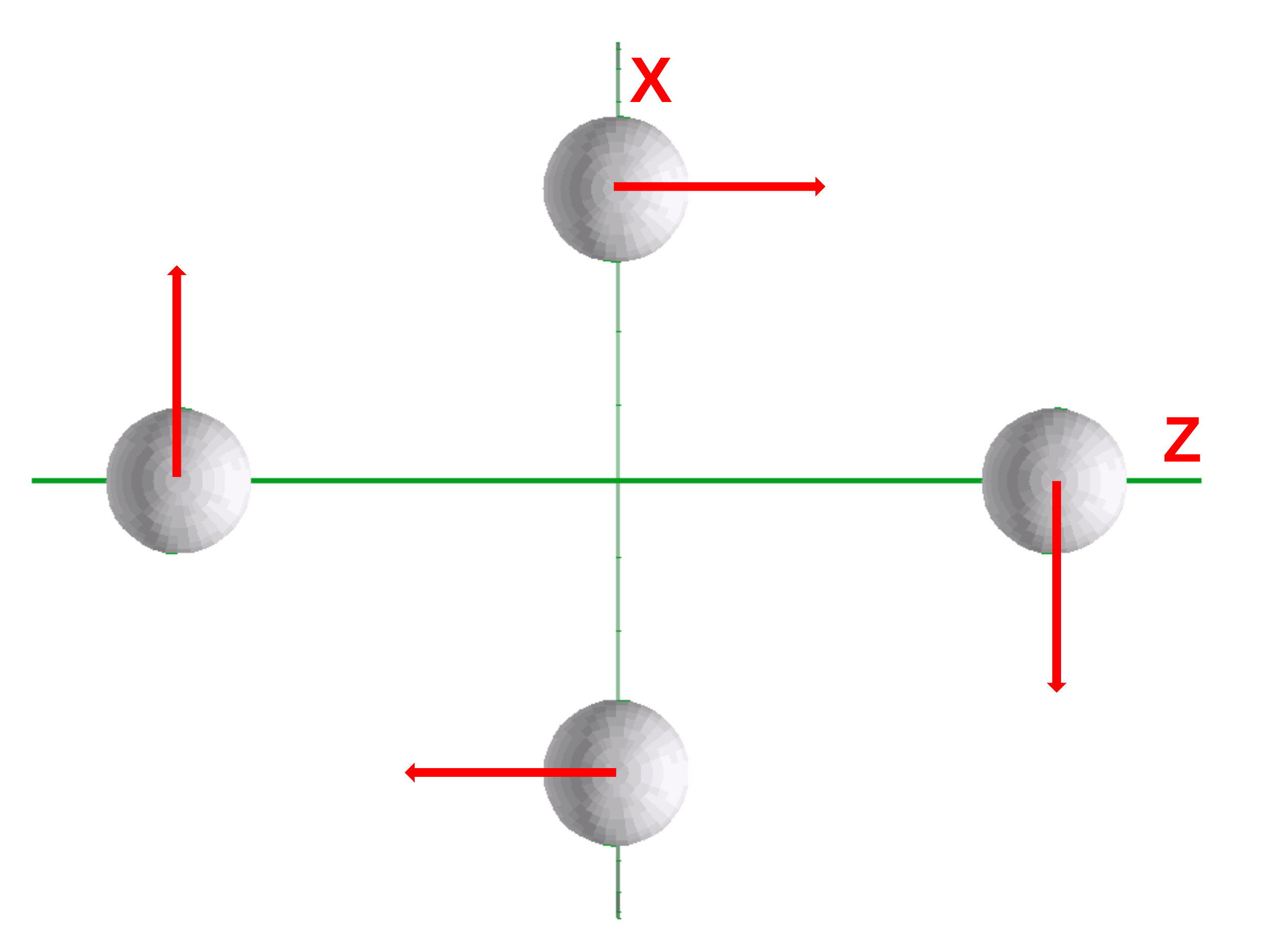}
\end{center}
\vskip -4mm
\caption{ (color online)
Four moving sources in the reaction ($[x-z]$) plane,
one pair, $s1$, is separated in the $x-$ directions and the other, $s2$,
is in the $z-$ direction. The sources are moving in the
directions indicated by the (red) arrows, $\pm \vec u_{s1}$ for the 1st pair
and $\pm \vec u_{s2}$ for the other.}
\label{F-9}
\end{figure}

It is important to mention that to detect rotation the accurate
identification of the reaction plane and its proper orientation 
is necessary. In the so called "cumulative" methods the reaction plane
is identified but its projectile and target sides are not.  This
makes it impossible to detect directed flow, and odd components
of the global collective flow. (All harmonic components
of random fluctuations of course can be detected.)  Furthermore,
not only the reaction plane with proper direction but also
the event by event center of mass (c.m.) should also be identified
\cite{Eyyubova}. This hardly ever done!  In both cases the use of
zero degree calorimeters are provide an adequate tool as these
are sensitive to the spectator residues.

The correlation function depends both on vectors $\vec k$ and $\vec q$.
To detect rotation the choices should be correlated correctly with
the beam and the directed reaction plane as illustrated in Fig.
\ref{F-8}. The positive x-axis points in the direction
of the projectile, which moves in the positive direction along the 
z-axis.

In Eq. (\ref{C2ms}), in the above situation, 
$\vec k \vec u_s = \gamma k_z v_z$, \ \ 
$\vec q \vec u_s = \gamma q_z v_z$ \ \ and  
$\vec q \vec x_s =  q_x d$. Thus, the Correlation function, apart of the 
single cell source size, $R$, sensitivity, has a specific dependence 
on $k_z$ and $q_z$, as well as on $q_x$. Unfortunately it is difficult 
to measure the particle momenta in the z-direction as it coincides with 
the beam. The $q_x$ dependence would enable us to estimate the
distance of the two sources.
\bigskip

%%%%%%%%%%%%%%%%%%%%%%%%%%%%%%%%%%%
\subsection{Four Fluid Cell Sources}
%%%%%%%%%%%%%%%%%%%%%%%%%%%%%%%%%%%

Four sources can be treated as a combination
of two moving double source systems. We use the same parameters 
as under paragraph \ref{Ltsm}, where $s_1$ and $s_2$ will be the two 
different pairs of sources with different locations and velocities.
See Fig. \ref{F-9}.

So we have
\begin{widetext} %____________________________________________________
\be
\begin{split}
C(k,q) & = 1 + \exp(-R^2 q^2) \times  \\
& \left[
\cosh\left(\frac{2 \vec k \cdot \vec u_{s_1}}{T_s} \right)+ 
\cosh\left(\frac{\vec q \cdot \vec u_{s_1}}{T_s} \right)
\cos(2\vec q \cdot \vec x_{s_1})+ 
\cosh\left(\frac{2 \vec k \cdot \vec u_{s_2}}{T_s} \right)+ 
\cosh\left(\frac{\vec q \cdot \vec u_{s_2}}{T_s} \right)
\cos(2\vec q \cdot \vec x_{s_2})
+  \right. \\
& 2\cosh\left(\frac{\vec k \cdot (\vec u_{s_1} - \vec u_{s_2})}{T_s}\right)
\cosh\left(\frac{\vec q \cdot (\vec u_{s_1} + \vec u_{s_2})}{2 T_s}\right)
\cos(\vec q \cdot (\vec x_{s_1} + \vec x_{s_2}))  +  \\
& \left.
2\cosh\left(\frac{\vec k \cdot (\vec u_{s_1} + \vec u_{s_2})}{T_s}\right)
\cosh\left(\frac{\vec q \cdot (\vec u_{s_1} - \vec u_{s_2})}{2 T_s}\right)
\cos(\vec q \cdot (\vec x_{s_1} - \vec x_{s_2}))
\right] \times  \\
& \left[
\cosh\left(\frac{2 \vec k \cdot \vec u_{s_1}}{T_s} \right)+ 
\cosh\left(\frac{2 \vec k \cdot \vec u_{s_2}}{T_s} \right) +
2\cosh\left(\frac{ \vec k \cdot (\vec u_{s_1} + \vec u_{s_2})}{T_s} \right) +
2\cosh\left(\frac{ \vec k \cdot (\vec u_{s_1} - \vec u_{s_2})}{T_s} \right) +
2  \right]^{-1} \\
\end{split}
\label{C4ms1}
\ee
\end{widetext} %________________________________________________
If $s_1 = s_2$ then we recover Eq. (\ref{C2ms}) 

In the case of a rotating but symmetric system the displacements
and velocities are of equal magnitude and are 
orthogonal to each other in the two pairs:
$\vec x_{s_1} \perp \vec x_{s_2}$ and
$\vec u_{s_1} \perp \vec u_{s_2}$. Thus a simple sign change of the 
velocity for one of the pairs or both does not change the result,
and so the rotation can be identified, 
but this evaluation does
not provide sensitivity to the direction of the rotation.
The reason is in the simplified freeze out assumption as we mentioned
already at the end of paragraph \ref{Ltsm}.

If the two pairs are not completely identical, i.e. the magnitude 
of the characteristic quantities of the two source pairs are not equal
then a sensitivity to the direction of the rotation may in principle
occur. However, if we change the direction of the velocities
of the two source pairs simultaneously (as it happens in
changing the direction of rotation) the result still does not 
change.

%%%%%%%%%%%%%%%%%%%%%%%%%%%%%%%%%%%
{\bf Four Sources with Flow Circulation:}
Recent fluid dynamical studies indicate
\cite{hydro1,hydro2}, that due to the initial
shear and angular momentum the early fluid dynamical development
has significant flow vorticity and circulation on the reaction
plane. These were recently evaluated \cite{CMW12}. At the present 
LHC Pb+Pb collision energy in the mentioned fluid dynamical model
calculation the maximum value of vorticity, $\omega$, was found 
exceeding $3$ c/fm , and the circulation after $6$ fm/c flow
development and expansion was still around 4-5 fm$\cdot$c. 
This vorticity in the reaction plane was more than an order of 
magnitude bigger than in the transverse
plane estimated from random fluctuations in ref. \cite{FW11}.

\begin{figure}[ht] %%%%%%%%%%
\begin{center}
      \includegraphics[width=7.6cm]{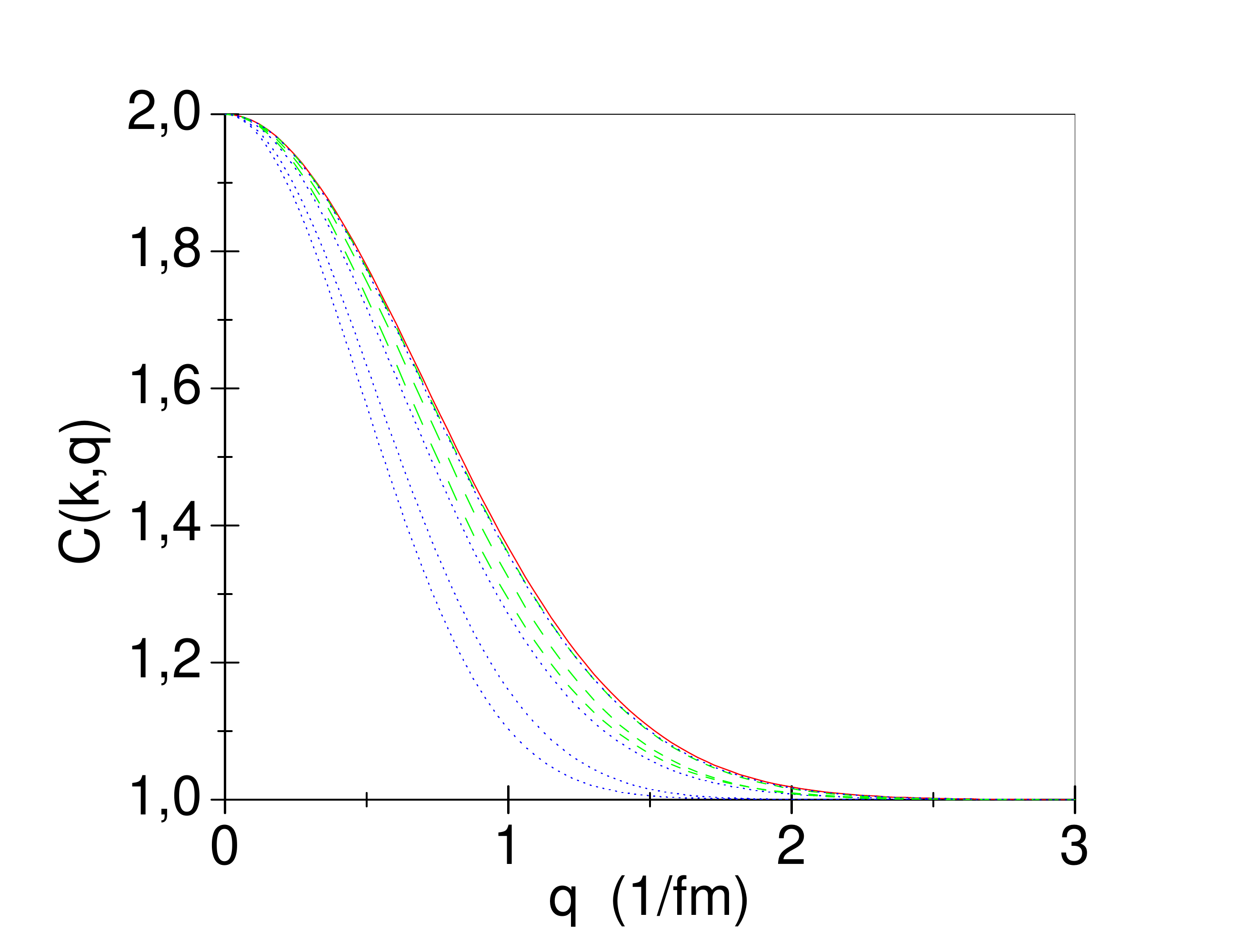}
\end{center}
%\vskip -6mm
\caption{ (color online)
The correlation functions, $C(k,q)$, for 4 sources 
where the displacement is such that there is one 
source pair on the x-axis and one the z-axis, 
the center-of-mass momentum of the emitted particles, 
$\vec k$, is in the x-direction. 
The {\it dotted blue lines} are for the velocity $v=0.5$\,c 
and displacement $d_x = d_z = 1.6$ fm. 
The {\it dashed green lines} are for the velocity $v=0.8$\,c 
and displacement $d_x = d_z = 1.0$ fm. $T_s = 0.20$ GeV.
In both cases the circulation is $\Gamma = 5$ fm$\cdot$c.
The values of $k_x$ are for the {\it blue lines}: 
0.5, 1.5, 3.0 and 6.0 fm$^{-1}$ and for the {\it green lines}: 
0.5, 1.5 and 3.0 fm$^{-1}$.
The {\it solid red line} is the correlation function $C(k_x,q_y)$. 
For large values of the center-of-mass momentum $k_x$ the 
correlation functions $C(k_x,q_x)$ and $C(k_x,q_z)$ 
will approach the correlation function $C(k_x,q_y)$. 
The larger displacement and smaller rotation velocity
leads to stronger deviation from the unaffected
correlation function $C(k_x,q_y)$.  }
\label{F-10}
\end{figure}

In this section we will look at the four source 
correlation function with similar circulation as in the above
mentioned fluid dynamical model estimates in the reaction plane.
See Fig. \ref{F-9}.
We will simulate a circulation value $\Gamma = 5 fm \cdot c$.
We use Eq. (\ref{C4ms1}) where the 
center-of-mass momentum, $\vec k$ points in the $x-direction$.

Since the position and velocity are of the same 
value and because of symmetry the correlation functions 
$C(k_x,q_x)$ and $C(k_x,q_z)$ provide the same values. 
So we take the correlation function $C(k_x,q_x)$ and 
we have afterwards some simplifications. See Fig. \ref{F-10}.

\be
\begin{split}
& C(k_x,q_x) = 1 + \exp(-R^2 q^2) \times \\
&
\left[ 1+\cos(2 q_x d) + \cosh\left(2\frac{k_x \gamma v_x}{T_s}\right) + 
\cosh\left(\frac{q_x \gamma v_x}{T_s}\right) + \right. \\
&
\left. 4 \cosh\left(\frac{k_x \gamma v_x}{T_s}\right) 
\cosh\left(\frac{q_x \gamma v_x}{2 T_s}\right) \cos(q_x d) \right]   \times \\
&
\left[\cosh\left(\frac{2 k_x \gamma v_x}{T_s}\right) + 
4 \cosh\left(\frac{k_x \gamma v_x}{T_s}\right) + 3 \right]^{-1}
\end{split}
\ee
\begin{figure}[ht] %%%%%%%%%%
\begin{center}
      \includegraphics[width=7.6cm]{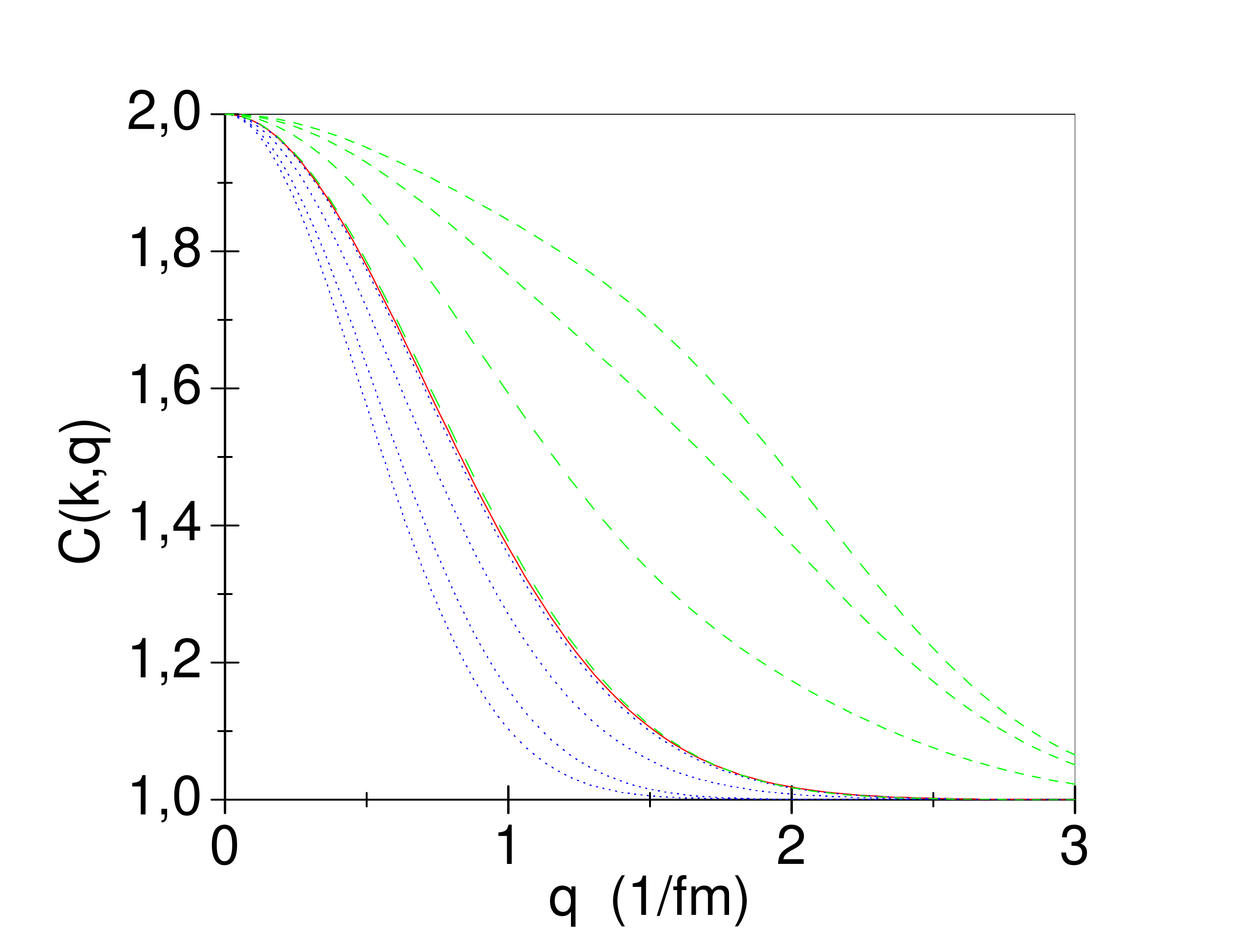}
\end{center}
\vskip -6mm
\caption{ (color online)
The correlation functions, $C(k,q)$, for 4 sources 
where the displacement is such that there is one 
source pair on the x-axis and one the z-axis, 
the center-of-mass momentum of the emitted particles, 
$\vec k$, is in the x-direction. 
The {\it dotted blue lines} are for the velocity $v=0.5$\,c 
and displacement $d_x = d_z = 1.6$ fm (same as in the previous figure). 
The {\it dashed green lines} are for the velocity $v=0.95$\,c 
and displacement $d_x = d_z = 0.84$ fm, $T_s = 0.20$ GeV.
The values of $k_x$ are: for the {\it blue lines} 
0.5, 1.5, 3.0 and 6.0 fm$^{-1}$ and for the {\it green lines} 
0.1, 0.25, 0.5 and 1.5 fm$^{-1}$.
The {\it solid red line} is the correlation function $C(k_x,q_y)$. 
For large values of the center-of-mass momentum $k_x$ the 
correlation functions $C(k_x,q_x)$ and $C(k_x,q_z)$ 
will approach the correlation function $C(k_x,q_y)$. 
Now for the dashed green lines with even higher velocity and 
smaller displacement, the deviation is significant and it is in the 
positive direction.
}
\label{F-11}
\end{figure}
For $C(k_x,q_y)$ we have the same result as 
we had for the two moving sources. Here the flow and displacement have no
effect.

Let us look at comparisons for similar circulations and 
for similar displacements. 

\begin{figure}[ht] %%%%%%%%%%
\begin{center}
      \includegraphics[width=7.6cm]{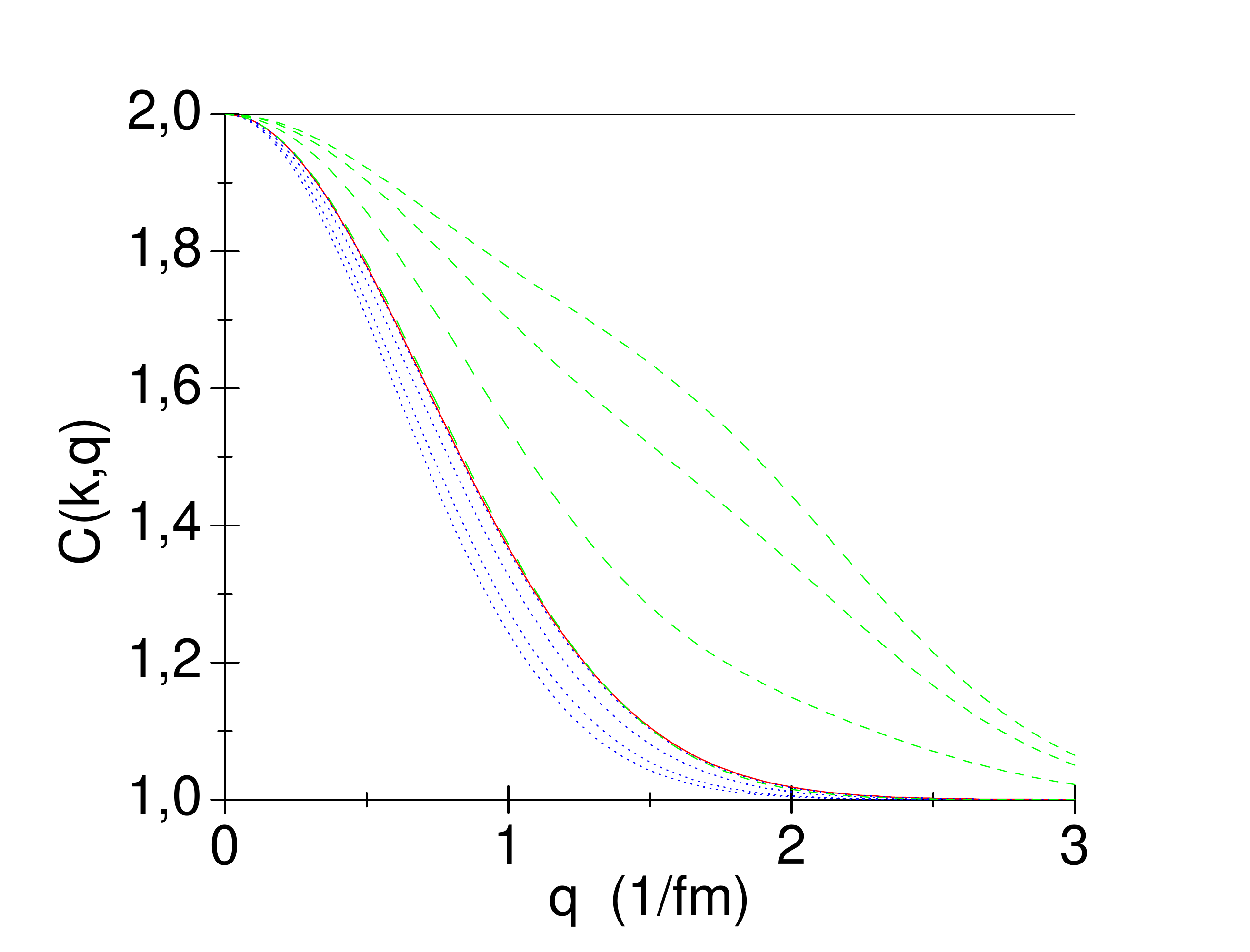}
\end{center}
\vskip -6mm
\caption{ (color online)
Same figure as the previous one, however the circulation 
is not the same and the displacements are equal.
The dotted blue lines are for the velocity $v=0.5$\,c 
and displacement $d_x = d_z = 1.0$ fm. 
The dashed green lines are for the velocity $v=0.95$\,c 
and displacement $d_x = d_z = 1.0$ fm. $T_s = 0.20$ GeV.
The values of $k_x$ are for the blue lines: 
0.5, 1.5, 3.0 and 6.0 fm$^{-1}$ and for the green lines: 
0.1, 0.25, 0.5 and 1.5 fm$^{-1}$.
Here the displacement is the same but the ultra-relativistic velocities
lead still to the deviation in the positive direction.}
\label{F-12}
\end{figure}
By comparing Figs. \ref{F-11} and \ref{F-12} we see that an 
increase in the displacement of the sources gives a increase in 
the apparent size of the system (narrower $q-$distribution. 
We also see that the measured size of the system increases with 
decreasing velocity.
At the same time the shape of correlation functions are becoming 
less and less Gaussian as the flow velocities increase. At the same time
the structure of the correlation function is also very different in 
different directions, which is not the case for spherical 
or linear expansion. This indicates that the rotating 
system contributes to essential non-Gaussian modifications,
which can be seen directly in the correlation function, but they would
become invisible if we would like to fit these data with a set of
Gaussians. Earlier works studied the correlation function at different 
angles or pseudorapidities with Gaussian parametrizations
\cite{DM95,Sin89},
however, for rotating systems this is not the most sensitive way of
presenting the results.
Thus rotation can be detected even in "symmetric" few source 
systems where the emission is equally probable from all emitting sources.
This emission scenario is less applicable to emission from
heavy ion reactions where the absorption of particles in QGP is 
not negligible, and this affects the emission from the interior of a
timelike (spacelike) FO layer, where the emission of earlier (deeper)
emitted particles are quenched. 

The correlation function is symmetric in all these cases as sources
from opposite sides of the system contribute equally. Thus the
correlation function is not sensitive to the direction of rotation.

%%%%%%%%%%%%%%%%%%%%%%%%%%%%%%%%%%%
\section{Asymmetric Sources}
%%%%%%%%%%%%%%%%%%%%%%%%%%%%%%%%%%%
\label{ASm}
\vskip -3mm

We have seen in the previous few source model examples that
a highly symmetric source may result in highly symmetric
correlation functions, however, this results were not sensitive 
to the direction of the rotation, which seems to be unrealistic.
We saw that this result is a consequence of the assumption that
both of the members of a symmetric pair contribute equally to the
correlation function even if one is at the side of the system
facing the detector and the other is on the opposite side. The
dense and hot nuclear matter or the Quark-gluon Plasma are
strongly interacting, and for the most of the observed particle types
the detection of a particle from the side of the system, -- which
is not facing the detector but points to the opposite direction, --
is significantly less probable. The reason is partly in the diverging 
velocities during the expansion and partly to the lower emission
probability from earlier (deeper) layers of the source from the
external edge of the timelike (or spacelike) FO layer. 
This feature is recognized for
a long time and discussed in detail by now. This influences
the particle emission (or freeze out (FO)) process and modifies
the post FO particle distribution. This topic has an extended
literature, and this feature destructs the symmetry of emission
of from source pairs at the opposite sides of the system
\cite{Sin89, M-2, M-3, M-4, CF, Si89, Bugaev, ALM99, ACG99, 
MAC99, Cs02, MAA03, TC04, MCM05, MCM06b}.

For the study of realistic systems where the emission is dominated
by the side of the system, which is facing the detector, 
we cannot use the assumption of the symmetry among pairs or 
groups of the sources from opposite sides of the system. Even if the
FO layer has a time-like normal direction, $\hat\sigma^\mu$ the 
$(k^\mu \hat\sigma^\mu)$ factor yields a substantial emission
difference between the opposite sides of the system.
Now we want to demonstrate this effect on few source examples,
and we will demonstrate the consequences of the non-symmetric emission.

%%%%%%%%%%%%%%%%%%%%%%%%%
\subsection{The Emission Probability}\label{TEP}
%%%%%%%%%%%%%%%%%%%%%%%%%

It was first recognized that the freeze out with the
Cooper-Fry description \cite{CF}, may lead to negative
contributions for particles, which move towards the center of 
the system and not in the direction out, towards the detectors.
The first proposal to remedy this problem came from Bugaev
\cite{Bugaev}, which led to the introduction of an improved
post freeze out distribution in the Cooper-Frye description,
first with the Cut-J\"uttner distribution \cite{Bugaev,ACG99}
and then by the Cancelling-J\"uttner distribution \cite{TC04}.

Subsequently it was realized that for the realistic
treatment of the freeze out process in transport theory
one has to modify the Boltzmann transport equation by
replacing the local molecular chaos assumption with
a non-local one, where the point of origin is also
included in the phase space distributions of the colliding
particles. This led to the Modified Boltzmann Transport
equation (MBT), and also the necessity to introduce
an escape probability, $P_{esc}$ was pointed out.

The escape probability was then introduced and analysed
in a series of publications \cite{M-2,M-3,M-4,MCM06b},
in transport theoretical approaches. It was pointed out that
even if the pre FO distribution is a locally equilibrated
isotropic distribution, the freeze out process and the
escape probability will provide a nonisotropic distribution
which eliminates the earlier observed problems. This
developing anisotropy in the freeze out process occurs
for freeze out both in space-like and time-like directions.

The escape probability introduced in the works 
\cite{M-2,M-3,M-4,MCM06b}, for a space-time surface layer 
of the system of thickness $L$, pointing in the four
direction $\hat\sigma^\mu$ was given at a point $x^\mu$ inside the
freeze out layer as
\be
P_{esc}(x) \ \propto \ 
\left(\frac{L}{L-x^\mu \hat\sigma_\mu}\right)
\left(\frac{p^\mu \hat\sigma_\mu}{ p^\mu u_\mu}\right)
\Theta(p^\mu \hat\sigma_\mu) \,,
\ee
where $p^\mu$ is the 
momentum of the escaping particle, $u^\mu(x)$ is the
local flow velocity and $s=x^\mu \hat\sigma_\mu$ is the distance of the
emission point from the inside boundary of the layer. The first
multiplicative term describes higher emission probability
to the particles, which are emitted closer to the outside boundary
of the layer, the second multiplicative term describes the
higher emission probability for the particles, which move in the
normal direction of the surface, because these should cross
less material in the layer. The last term secures that only those 
particles can escape, which move outwards through the layer.

The last two momentum dependent factors are important in
transport theoretical models, to determine the shape of 
the post FO momentum distribution, e.g. \cite{TC04}, 
which would replace the J\"uttner distribution.
This shape modification happens to the single and two
particle distributions equally, and it acts in all
emission directions, $\vec k$, equally, so this effect is
secondary from the point of view of the flow velocity
dependence of the correlation function.

In order to describe the complete freeze out process for
a reaction the system had to be surrounded with a freeze
out layer in the space-time, and the phase space 
distribution of the escaping, frozen out particles
can be obtained by integrating over the whole
4-volume of the freeze out layer the local (usually
isotropic) phase space distribution with the
escape probability $P_{esc}(x)$. This procedure 
would then play the role of function $G(x)$ in the
source function in Eq. (\ref{S-hydro}) instead of the
simplified assumptions, as e.g. in Eq. (\ref{G1}).

The correlation function, $C(k,q)$ is always measured
in a given direction of the detector, $\vec k$. Obviously
only those particles can reach the detector, which 
satisfy   $ \ k^\mu \hat\sigma_\mu \ > \ 0 $. Thus 
in the calculation of  $C(k,q)$ for a given $\hat{\vec k}$-
direction we can exclude the parts of the freeze out
layer where $ \ k^\mu \hat\sigma_\mu \ < \ 0 $ (see Eq. (10)
of ref. \cite{Sin89} or ref. \cite{Bugaev}). For time-like 
FO a simplest approximation for the emission possibility is
$
P_{esc}(x) \ \propto \  k^\mu u_\mu(x) 
$
\cite{Cso-5}. 

In a model calculation we therefore have to define the 
freeze out layer also, this realistically should not 
include the whole space-time volume of the reaction.
In case of calculating $C(k,q)$ for a given $\hat{\vec k}$
we should select the relevant part of the freeze
out layer, which may contribute to emission in the
$\vec k$ direction. This should be a layer of
2-3 m.f.p facing the detector at the direction
$\hat{\vec k}$. This can eliminate the symmetric pairs
of fluid cells in the previous calculations of the 
correlation function, even if the emission normal is timelike,
because the FO particle from an earlier emission point
in the ST has to propagate through the plasma for some
finite time, with considerable quenching.

Therefore in the following models we should apply the
escape probability and we should define a 
$\hat{\vec k}$-dependent freeze out layer also!
The most simple approximation is to select an emission
layer from the system for a given $\hat{\vec k}$-direction
with uniform emission probability from within this layer.
The next to most simple approximation is to introduce
an emission probability within the layer, increasing
towards the outside boundary of the layer. (Here it is 
important to mention that the spatial emission probability
should be sufficiently smooth, so that one fluid cell
and its contribution to $C(k,q)$, should not be effected
by this emission probability.

When we have up to 4 sources we can always add 
$\hat{\vec k}$-dependent emission weights to these 
sources. This still would qualitatively change the 
outcome. As we discuss here up to four sources only
a detailed formal evaluation of the emission probability
would be an exaggerated approach, by defining more
parameters than the outcome, so we just define the weights
themselves here. In a full 3+1D fluid dynamical model with
100000+ fluid cells of course we have to apply a realistic
and general evaluation of emission probability for every
point of the ST.

%%%%%%%%%%%%%%%%%%%%%%%%%
\subsection{Emission probabilities for few sources}\label{EPFS}
%%%%%%%%%%%%%%%%%%%%%%%%%

%%%%%%%%%%%%%%%%%%%%%%%%%
{\bf Two sources:}
The previous discussion included two sources (i) in the
beam-, $z-$direction and (ii) in the transverse direction in the
reaction plane, $x-$direction. In case (i) the emission 
could be different from the two sources if the detector
is in the $z-$ direction, which is difficult to achieve, so we
do not have to discuss this possibility.

\begin{figure}[ht] %%%%%%%%%%
\begin{center}
      \includegraphics[width=6cm]{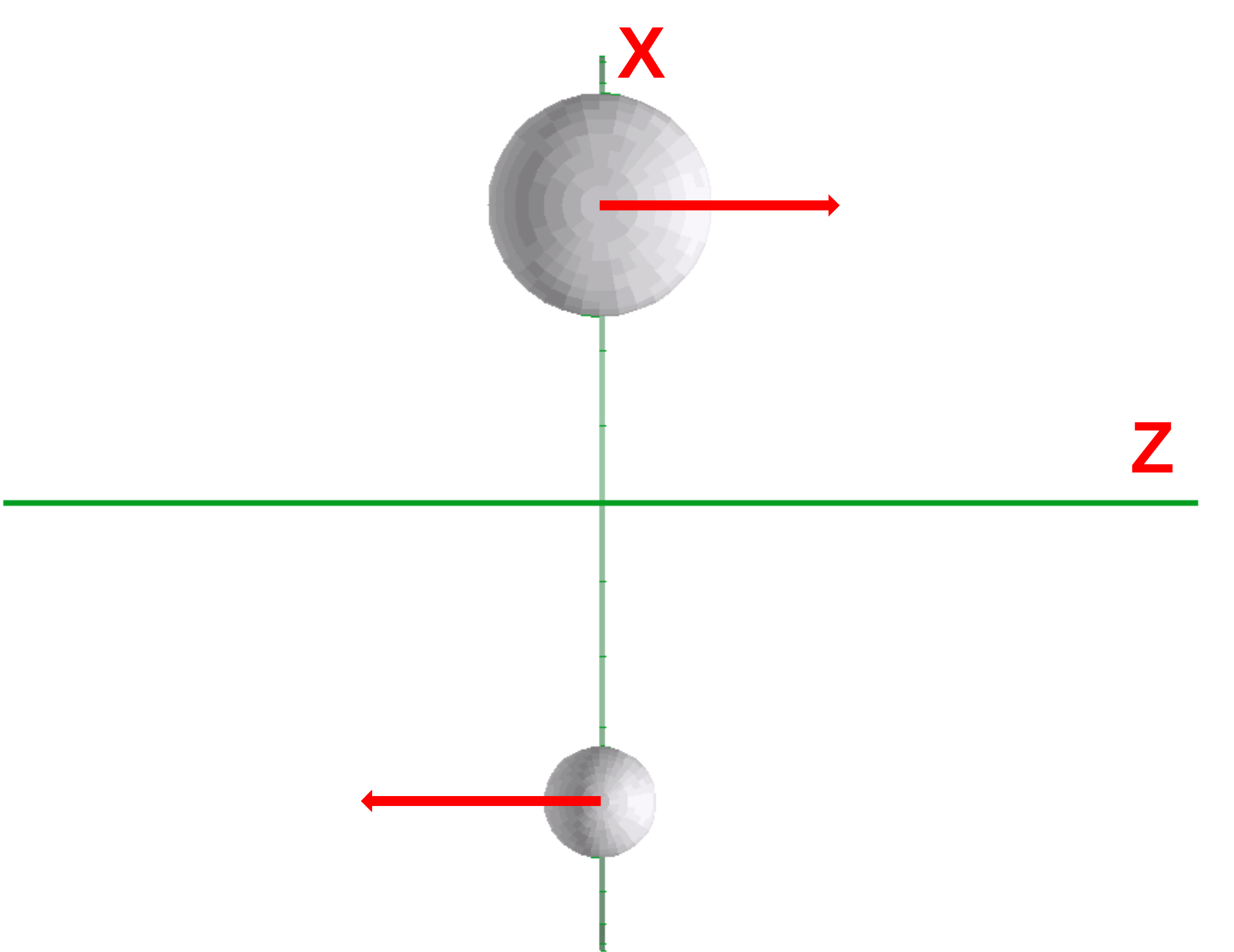}
\end{center}
\vskip -6mm
\caption{ (color online)
Two moving sources in the reaction ($[x-z]$) plane,
separated in the $x-$ direction (case (ii) in the text). 
The sources are moving in the
directions indicated by the (red) arrows. The detector is in the 
positive $x-$direction, thus the source on this side has more
dominant emission into this direction, and this is indicated 
by the bigger size of the source on this side.}
\label{F-13}
\end{figure}

In configuration (ii) the observation can be in
different $\hat{\vec k}$-directions. If $\hat{\vec k}$ points
into the $\pm y-$direction, then the probabilities must
be identical so emission probabilities do not lead to any change.

If $\hat{\vec k}$ points into the $\pm x-$direction,
then one of the sources is closer to the detector and may
shadow the more distant one. Thus, we can just introduce two
positive weight factors so that $w_c$ is the weight for the cells closer
to the detector and $w_s$ is for the cells which are far from
the detector measuring the average momentum $\vec k$. These
weights are the same for the calculation of the nominator and denominator
of the correlation function, so their normalization does not
influence the correlation function. 

As not all emitted
particles reach a given detector the normalization is also dependent
on the direction of the detector. Thus, we evaluate the 
correlation function this way. This immediately changes the earlier
result (\ref{C2kx}), because it breaks the symmetry between the
two sources. We can simply repeat the calculation
for two moving sources in section \ref{Ltsm}, modifying the derivation 
of Eq. (\ref{C2msd})
and obtain the general result
\be
\begin{split}
& \ \ \ \ \ \ \left. C(k,q)\!\!\phantom{\frac{1}{2}}\!\!\right|_{+x} 
= 1 + \exp(-R^2 q^2) \ \times\\ 
& 
\frac{
 w_c^2 e^{ \frac{2\vec k \vec u_s}{T_s}}  +
 w_s^2 e^{-\frac{2\vec k \vec u_s}{T_s}}  +
 2 w_c w_s \cosh\left( \frac{ \vec q \vec u_s}{T_s} \right) 
  \cos( 2\vec q \vec x_s)  
}{  
 w_c^2 e^{ \frac{2\vec k \vec u_s}{T_s}}  +
 w_s^2 e^{-\frac{2\vec k \vec u_s}{T_s}}  + 2 w_c w_s  } \ .
\end{split}
\label{C2msAs}
\ee

Note that this result is valid for the case when $\hat{\vec k}$ 
points to the $+x$ direction, because the weights depend on this
and $w_c > w_s$. See Fig. \ref{F-13}. The fact that the emission
from the source, which is closer to the detector is stronger
makes the direction of the flow detectable.
\bigskip

If we introduce the notation $w_c = 1+\epsilon$
and $w_s = 1-\epsilon$, the deviation from the symmetric
result will become apparent

\begin{widetext}  %----------------------------------------------------
\be
\left. C(k,q)\!\!\phantom{\frac{1}{2}}\!\!\right|_{+x} 
= 1 + \exp(-R^2 q^2) \ 
\frac{
 (1+\epsilon^2)  \cosh\left(\frac{2\vec k \vec u_s}{T_s}\right)  +
 2 \epsilon      \sinh\left(\frac{2\vec k \vec u_s}{T_s}\right)  +
 (1-\epsilon^2)  \cosh\left( \frac{ \vec q \vec u_s}{T_s} \right) 
  \cos( 2\vec q \vec x_s)  
}{  
 (1+\epsilon^2)  \cosh\left(\frac{2\vec k \vec u_s}{T_s}\right)  +
 2 \epsilon      \sinh\left(\frac{2\vec k \vec u_s}{T_s}\right)  +
 (1-\epsilon^2)  } \ .
\label{C2msAsw}
\ee
\end{widetext} %------------------------------------------------------

If $\epsilon \rightarrow 0$, i.e. if $w_c=w_s$, we recover the earlier
result, Eq. (\ref{C2ms}).

\begin{figure}[ht] %%%%%%%%%%
\begin{center}
      \includegraphics[width=6cm]{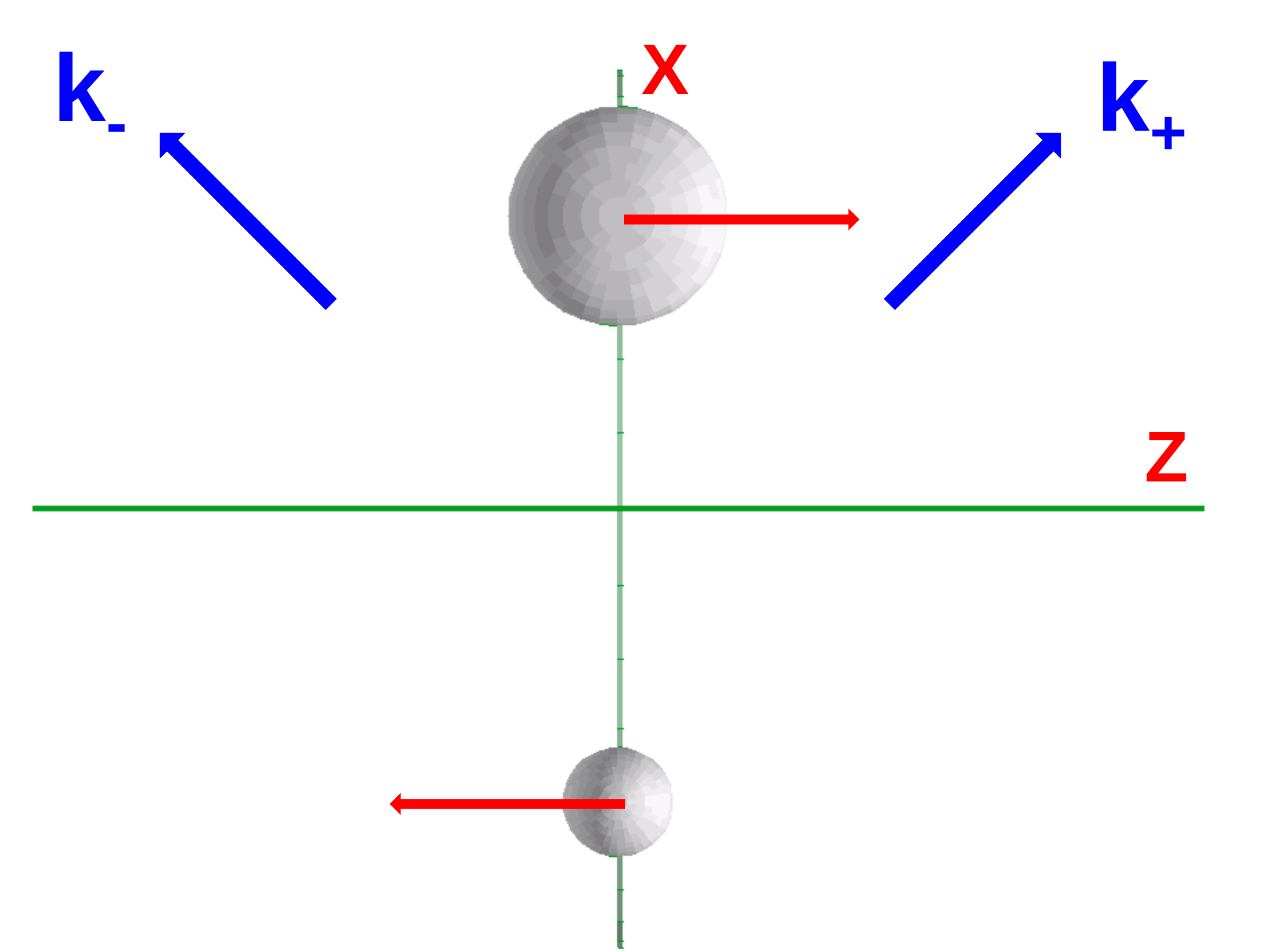}
\end{center}
\vskip -4mm
\caption{ (color online)
Two moving sources in the reaction ($[x-z]$) plane,
separated in the $x-$ direction. 
The sources are moving in the
directions indicated by the (red) arrows. The two "tilted" detector
directions are indicated by the (blue) arrows labeled with
$k_+$ and $k_-$.}
\label{F-14}
\end{figure}

If $\epsilon = 0$ we have the symmetric situation where both
sources have equal contribution, the asymmetric terms vanish, and
the result becomes to be symmetric for the change of the direction
of the flow velocity.  If $\epsilon$ reaches its maximal value,
$\epsilon = 1$ the contribution of the far side source is eliminated
($w_s=0$, $w_c=2$), and 
only the single nearby source contributes to the correlation function. In 
this case the asymmetric term in the nominator vanishes, the remaining 
terms in the nominator and denominator are equal, and we recover the
single static source result.
 
This result has terms, which change sign if the flow velocity, ${\vec u}_s$
changes sign. The result is valid only if the detector is in the 
$\hat{\vec k} = (1,0,0)$ direction. For this direction, however, if the
flow velocity points in the $z$-direction, i.e. orthogonal to $\vec k$
the asymmetric term does not provide any contribution, so
it will not show up in $C(k_x,\vec q)$. To circumvent this problem
we should study detector directions, which do not coincide with
the primary axes of the given event (where $x$ is the direction of 
the impact parameter vector, $\vec b$, pointing to the projectile;
$y$ is the other transverse direction; and $z$ is the direction of the 
projectile beam).

\begin{figure}[ht] %%%%%%%%%%
\begin{center}
      \includegraphics[width=7.9cm]{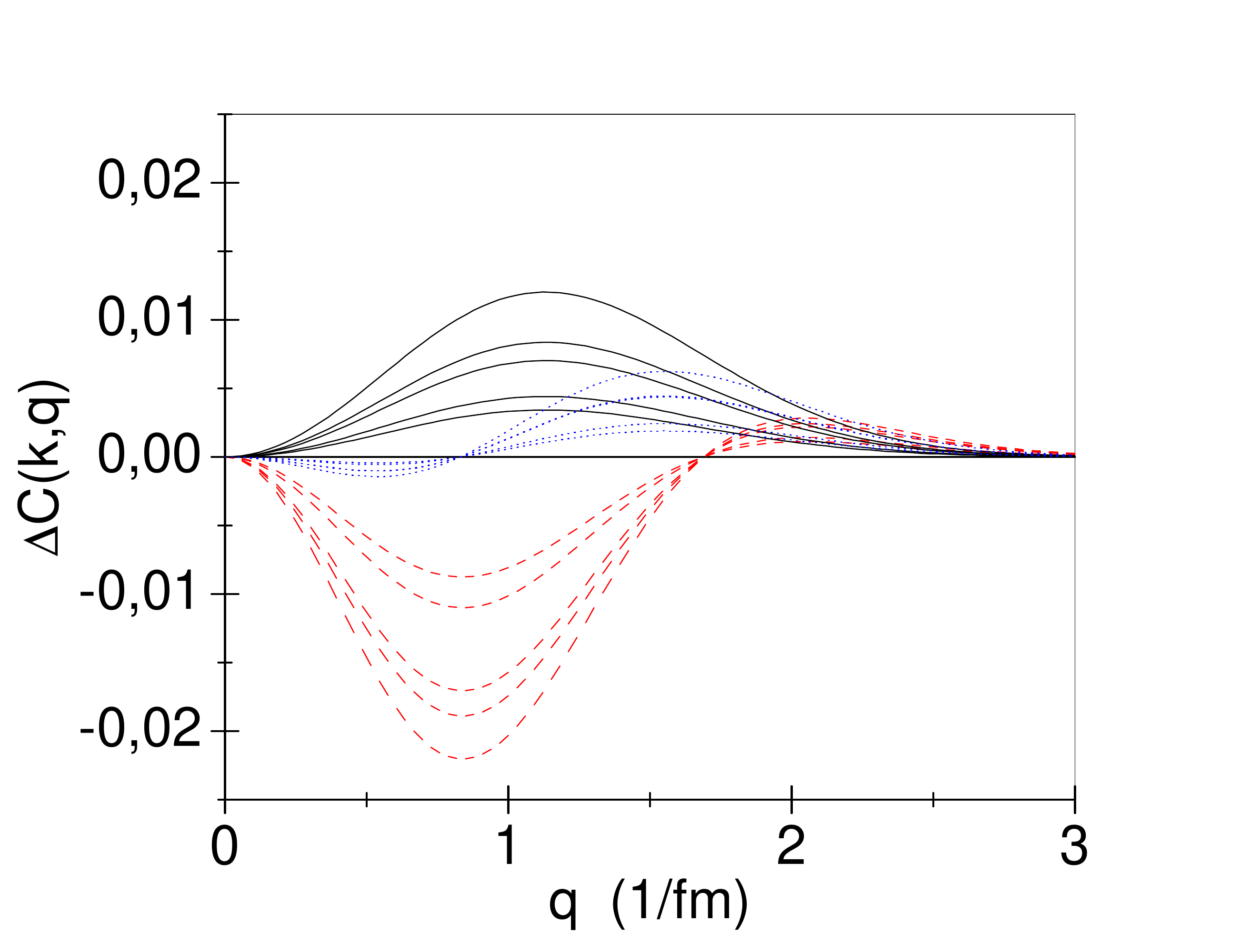}
\end{center}
\vskip -0.7cm
\caption{ (color online)
Difference of the forward and backward shifted 
correlation function,  $\Delta C(k_\pm,q_{out})$, 
for the value $\epsilon = 0.50$.
The {\it solid black lines} are for the velocity $v_z=0.5$\,c,
{\it dotted blue lines} are for the velocity $v_z=0.6$\,c and
{\it dashed red lines} are for the velocity $v_z=0.7$\,c.
Displacement is $d_x = 1.0$ fm, $T_s = 0.139$ GeV 
and $a = b = 1/\sqrt{2}$.
The values of $k$ are: for the {\it solid black lines} 
0.25, 0.50, 2.00, 2.75 and 3.50   fm$^{-1}$,
the {\it dotted blue lines}
0.25, 0.50, 1.00, 1.75 and 2.50 fm$^{-1}$,
and the {\it dashed red lines}
0.25, 0.50, 0.75, 1.25 and 1.75 fm$^{-1}$.}
\label{F-15}
\end{figure}

%%%%%%%%%%%%%%%%%%%%%%%%%%%%%%%%
{\bf Correlation in Tilted Directions:}
The form of the correlation function is the same if $\vec k$ is in the 
same plane, the reaction plane, 
but it has a $z$ component also, i.e. $\vec k = (k_x, 0, \pm k_z)$.
This is possible for all LHC heavy ion experiments, ATLAS, CMS and
even ALICE, where the longitudinal acceptance
range of the TPC ($\Delta \eta \ < 0.8$) is the smallest. 
See Fig. \ref{F-14}.

Earlier for spherically or longitudinally expanding systems the dependence
of the correlation function on the tilt angle or {\it width} parameter
was analysed in detail in ref. \cite{Sin89}. We do not go into similar
fine details, just demonstrate the possibilities for an arbitrary
configuration.

Depending on the detector 
acceptance we should chose a detector direction where $|k_z|$ is as big
as the detector acceptance allows it.    
For this configuration the {\bf form} of the correlation function is 
the same as (\ref{C2msAsw})
\be
\left. C(k,q)\!\!\phantom{\frac{1}{2}}\!\!\right|_{+x,\pm z} =
\left. C(k,q)\!\!\phantom{\frac{1}{2}}\!\!\right|_{+x} \ ,
\ee
with keeping the different weights, $w_c, w_s$ or $\epsilon$ so that
the forward shifted and backward shifted directions have the same weights. 
These weights are not specified up to now anyway.

For detection of the correlation function we have to introduce here
the usual, $\vec k$-dependent coordinate system to classify the
direction of $\vec q$.
Thus if 
\be 
\hat{\vec k}_\pm = (a,0,\pm b){\rm fm}^{-1} , \ 
k_x = a | \vec k | , \
k_z = \pm b | \vec k | , 
\label{kcomp}
\ee
where $a^2 + b^2 = 1$, see Fig. \ref{F-14}.
Then the difference vector, $\vec q$, can be measured in the directions
\be
\begin{split}
& \hat{\vec q}_{out}  = (a,0, \pm b) , \
q_x = a | \vec q | , \ q_z = \pm b | \vec q | \\
& \hat{\vec q}_{side} = (0,1,0) , \ \ \ q_y = | \vec q | \\
& \hat{\vec q}_{long} = (\mp b,0,a) , \ \ 
q_x = \mp b | \vec q | , \ q_z = a | \vec q | .
\end{split}
\label{qcomp}
\ee
This leads to the following correlation functions

\begin{widetext} %-------------------------------------------------------

\begin{equation}
\begin{split}
& C(k_{(\pm)},q_{out})  = 1 +  \exp(-R^2 q^2)\, 
\frac{
 (1{+}\epsilon^2) \cosh\left( \frac{2 \gamma k_z  v_z}{T_s} \right) +
 2 \epsilon       \sinh\left( \frac{2 \gamma k_z  v_z}{T_s} \right) +
 (1{-}\epsilon^2) \cosh\left( \frac{  \gamma q_z  v_z}{T_s} \right) 
                   \cos\left( q_x  d_x \right)   
}{  
 (1{+}\epsilon^2) \cosh\left( \frac{2 \gamma k_z  v_z}{T_s} \right) +
 2 \epsilon       \sinh\left( \frac{2 \gamma k_z  v_z}{T_s} \right) + 
  (1{-}\epsilon^2) } \ , \\
& C(k_{(\pm)},q_{side}) = 1 + \exp(-R^2 q^2)\ , \\
& C(k_{(\pm)},q_{long}) = 1 + \exp(-R^2 q^2)\,
\frac{
 (1{+}\epsilon^2) \cosh\left( \frac{2 \gamma k_z  v_z}{T_s} \right) +
 2 \epsilon       \sinh\left( \frac{2 \gamma k_z  v_z}{T_s} \right) +
 (1{-}\epsilon^2) \cosh\left( \frac{  \gamma q_z  v_z}{T_s} \right) 
                   \cos\left( q_x  d_x \right)   
}{  
 (1{+}\epsilon^2) \cosh\left( \frac{2 \gamma k_z  v_z}{T_s} \right) +
 2 \epsilon       \sinh\left( \frac{2 \gamma k_z  v_z}{T_s} \right) + 
  (1{-}\epsilon^2) } \ .
\end{split}
\label{C2kx1}
\end{equation}

Although, it seems that  $C(k_{(\pm)},q_{out})$ and $C(k_{(\pm)},q_{long})$
are the same, this is in fact not the case, because the values of the
components of the different types of $\vec k$ and $\vec q$ are not the same
as described in Eqs. (\ref{kcomp},\ref{qcomp}). In all cases,
the out-, side- and long- $ q = | \vec q |$. We will also use the notation
$k = |\vec k|$ and $\gamma v_x = u_x$, $\gamma v_y = u_y$, $\gamma v_z = u_z$,
so that $\vec u_s = (u_x, u_y, u_z)$.
For example for the {\it out}
component the difference of the forward and backward shifted correlation 
functions is

\begin{equation}
%\begin{split}
%& 
\Delta C(k_\pm, q_{out}) \equiv   
% \\
%
% & 
C(k_{+},q_{out}) -  C(k_{-},q_{out}) =
\frac{4 \exp(-R^2 q^2)\ 
  \epsilon  \sinh\left( \frac{2 u_z\, b k}{T_s} \right)\ 
  (1{-}\epsilon^2)\left[ 1 - 
   \cosh\left( \frac{ u_z\, b q}{T_s} \right) 
                   \cos\left(a q d_x \right) \right]
}{
 \left[(1{+}\epsilon^2) \cosh\left(\frac{2 u_z\, b k}{T_s}\right) 
      +(1{-}\epsilon^2) \right]^2 - 
  4 \epsilon^2  \sinh^2\left(\frac{2 u_z\, b k}{T_s} \right)}.
% \end{split}
\label{DCpm}
\end{equation}

\end{widetext}   %--------------------------------------------------------

As Eq. (\ref{DCpm}) and Fig. \ref{F-15} show, the Differential Correlation 
Function (DCF),  $\Delta C(k_\pm,q_{out})$, 
is sensitive to the speed and direction 
of the rotation, and it is also sensitive to the amount of the 
tilt in the directions of the detection, regulated here by the
parameters $a$ and $b$.
$\Delta C(k_\pm,q_{out})$ tends to zero both if $q \rightarrow 0$
and if $q \rightarrow \infty$. The structure of 
$\Delta C(k_\pm,q_{out})$ is determined by the 
$\cosh\left( u_z b\, q / T_s \right)\, \cos\left(a d_x\, q \right)$
product. If in both arguments the coefficients of $q$, 
$ u_z b / T_s$ and $a d_x$ are positive, smaller than one,
and $ u_z b / T_s \le a d_x$,
then the DCF is positive.
If the coefficient $a d_x$ exceeds one the $\cos$ 
function changes sign at high $q_{out}$ values (e.g.
above $q = 1-2$fm$^{-1}$),
and the DCF becomes 
negative at high $q_{out}$ values. Note that the
ratio of the two coefficients is influenced be the 
choice of the tilting angle, i.e. by the parameters
$a$ and $b$.

\begin{figure}[ht] %%%%%%%%%%
\begin{center}
      \includegraphics[width=6cm]{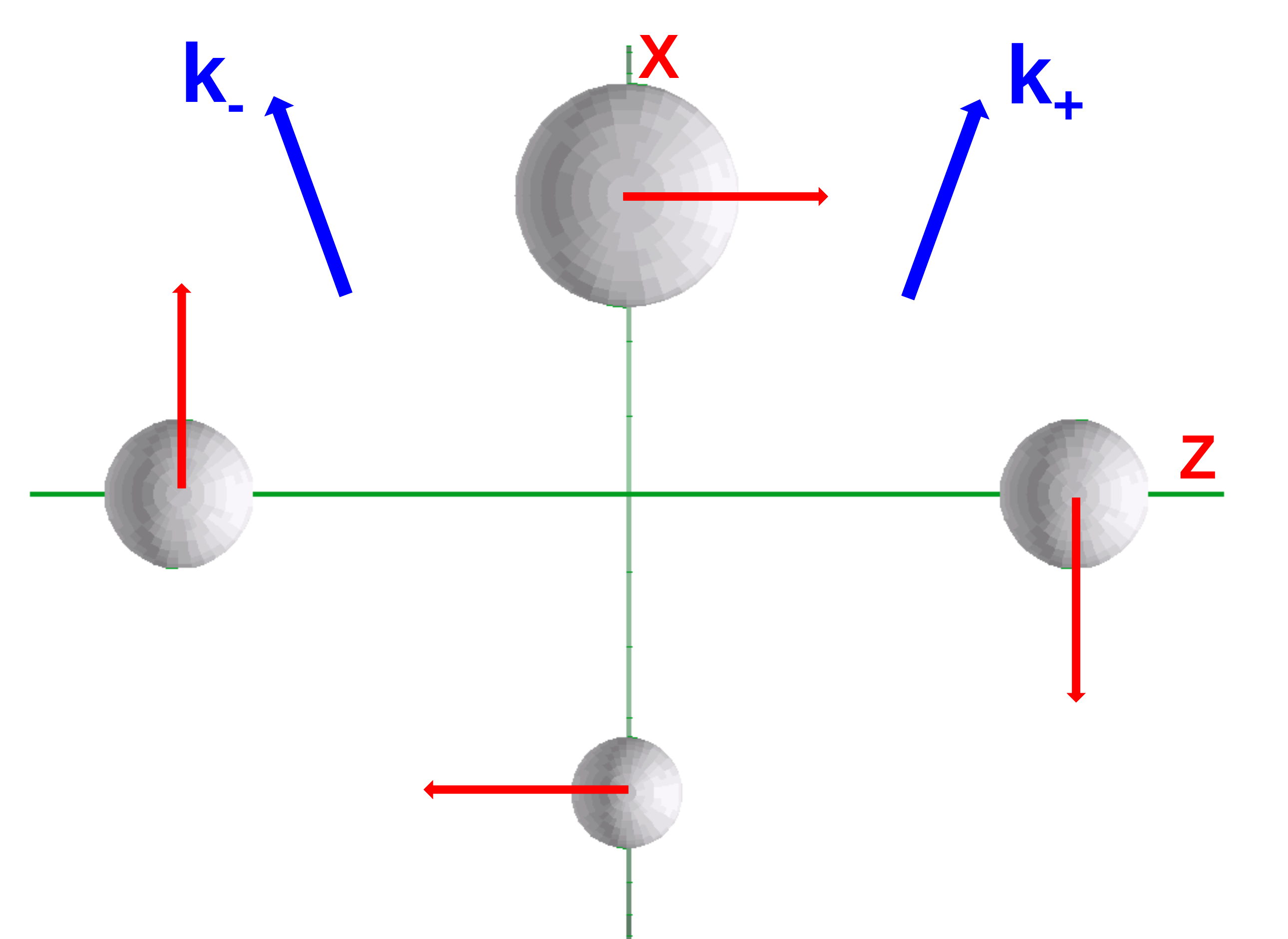}
\end{center}
\vskip -4mm
\caption{ (color online)
Four moving sources in the reaction ($[x-z]$) plane,
separaed in the $x-$ and $z-$ directions. 
The sources are moving in the
directions indicated by the (red) arrows. The "tilted" detector
directions are indicated by the (blue) arrows.}
\label{F-16}
\end{figure}

If the parameter $ a d_x $ remains constant,
about 1 fm, and then when $u_z b / T_s $ becomes larger (than one)
the Differential Correlation Function becomes negative
at small $q$ values.

If the parameter $ u_z b / T_s$ remains constant, and
about 1 fm, then when $a d_x$ becomes less (than one)
the DCF becomes negative
at small $q$ values.

In case if the detector has a narrow pseudorapidity acceptance, then
$\vec k_\pm$ is close to $k_x$, i.e. $b \ll a$ and then the weights 
are maximal for the source in the $x-$direction, as indicated in Fig.
\ref{F-16}.

If we change the direction of rotation to the opposite
the Differential Correlation Function changes sign due
to the $\sinh$ function in the nominator. In this configuration
with the change of the tilt of the detector directions
we can adjust the DCF,
to the threshold value where the $\Delta C(k_\pm,q_{out})$,
is still positive, which provides a sensitive estimate
for the rotation velocity at Freeze Out.

This very sensitive behaviour is rather special 
and it appears in this special two source model this
way.  With an increased resolution and with more
source elements this strong and specific structure
will be smoothed out to some extent.

\begin{figure}[ht] %%%%%%%%%%
\begin{center}
      \includegraphics[width=7.8cm]{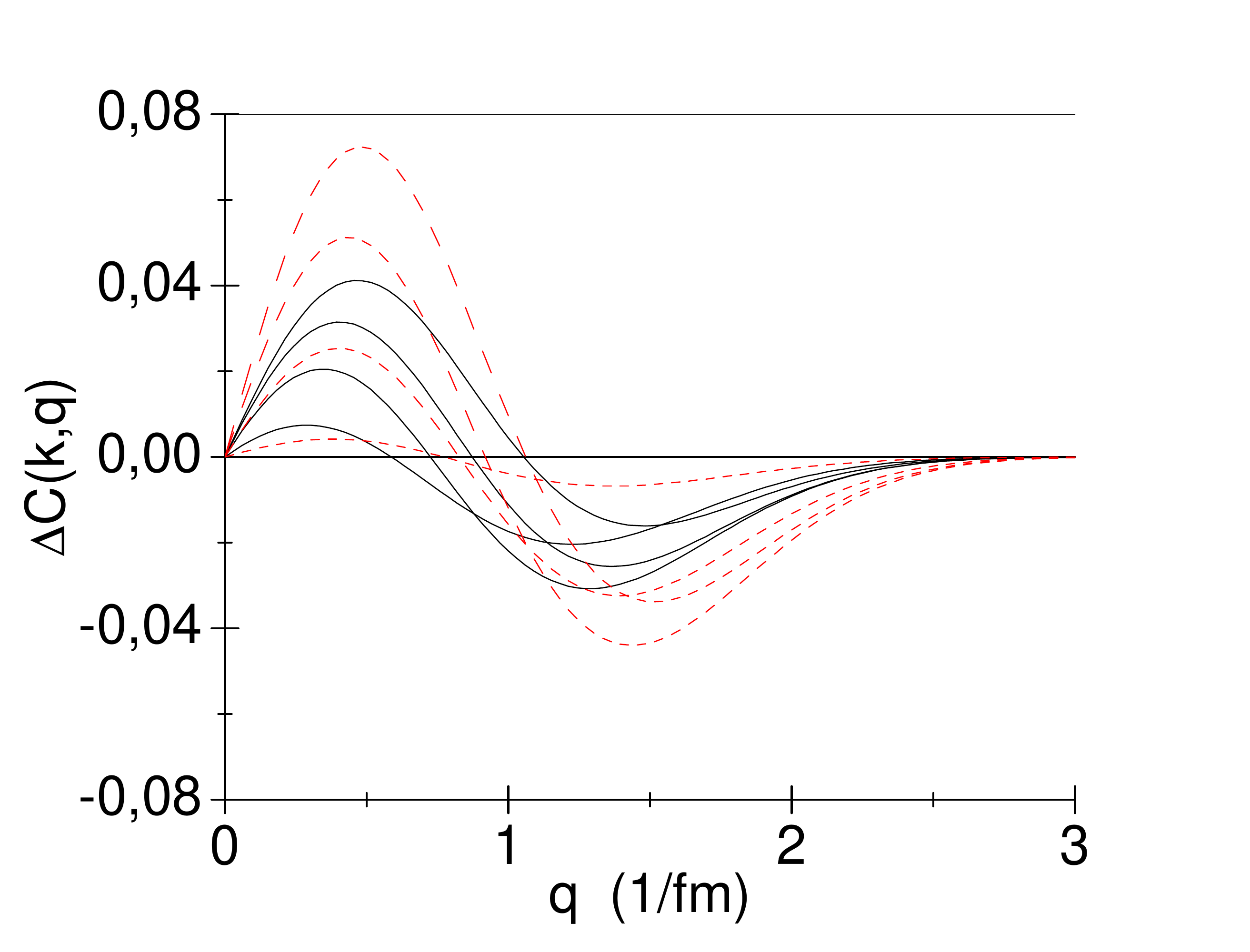}
\end{center}
\vskip -7mm
\caption{ (color online)
Correlation function difference for the weight factors:
$\omega_a = 1.25$, $\omega_b = 0.75$,
$\omega_c = \omega_d = 1.00$
for sources placed at $+x$, $-x$, $+z$ and $-z$ respectively. 
This weight distribution corresponds to the configuration
shown in Fig. \ref{F-16}.
The {\it solid black lines} are for the velocity $v_z=0.5$\,c,
and the {\it dashed red lines} are for the velocity $v_z=0.7$\,c.
Displacement is $d_x = d_z = 1.0$ fm, $T_s = 0.139$ GeV 
and $a = b = 1/\sqrt{2}$.
The values of $k$ are: for the {\it solid black lines} 
0.10, 0.50, 1.00, and 2.00 fm$^{-1}$ and
for the {\it dashed red lines}
0.10, 0.50, 1.00, and 2.00 fm$^{-1}$.
The difference is larger for smaller values of $k$.}
\label{F-17}
\end{figure}

The $\sinh( 2 u_z\, b k /T_s )$ term changes sign in the 
nominator when $u_z$ changes sign the difference of the two 
correlation functions, $\Delta C(k_\pm,q_{out})$ changes sign also
because all other terms are symmetric to the sign change of the 
velocity. 

This is an important observation as we can detect the
direction and magnitude of the rotation in the reaction plane.
This difference is also increasing with the longitudinal shift, $b$, 
of the average momentum vector, $\vec k$, so that detectors with
larger pseudorapidity acceptance can detect the rotation better. 

  In order to perform this measurement, one has to determine the
global reaction plane (e.g. from spectator residues in the ZDCs), and 
determine the projectile side of this plane. Furthermore
the event by event center of mass should also be identified
(using e.g. the method shown in ref. \cite{Eyyubova}). This will be the 
positive $x$-direction. Then the correlation function can be measured
for four different $\vec k$-directions in the global reaction plane.
These four directions are shifted forward and backward from the
center of mass symmetrically on the projectile side, and there should
be a symmetric pair of detection points in the target side of the
reaction plane too.

The $\vec k$ directions opposite to each other across the c.m. point
give the same result, while the difference, $\Delta C(k_\pm,q_{out})$, 
between the Forward (F) and 
Backward (B) shifted contributions will characterize the speed and direction
of the rotation. This symmetry can be used to eliminate the contribution
from eventual random fluctuations.  The observed F/B asymmetry depends on
the parameters $\epsilon$, $v_z$ and $d_x$, these can be estimated by
measuring the correlation functions at all possible moments $\vec k$.

Fig. \ref{F-15} indicates that the differential correlation function
has a larger amplitude for smaller $k$ values, and the zero points
are sensitively dependent on the rotation velocity.

The zero points come from the term
\be
1 - \cosh\left(\frac{u_z bq}{T_s}\right) \cos(a q d_x) = 0
\ee

\begin{figure}[ht] %%%%%%%%%%
\begin{center}
      \includegraphics[width=7.8cm]{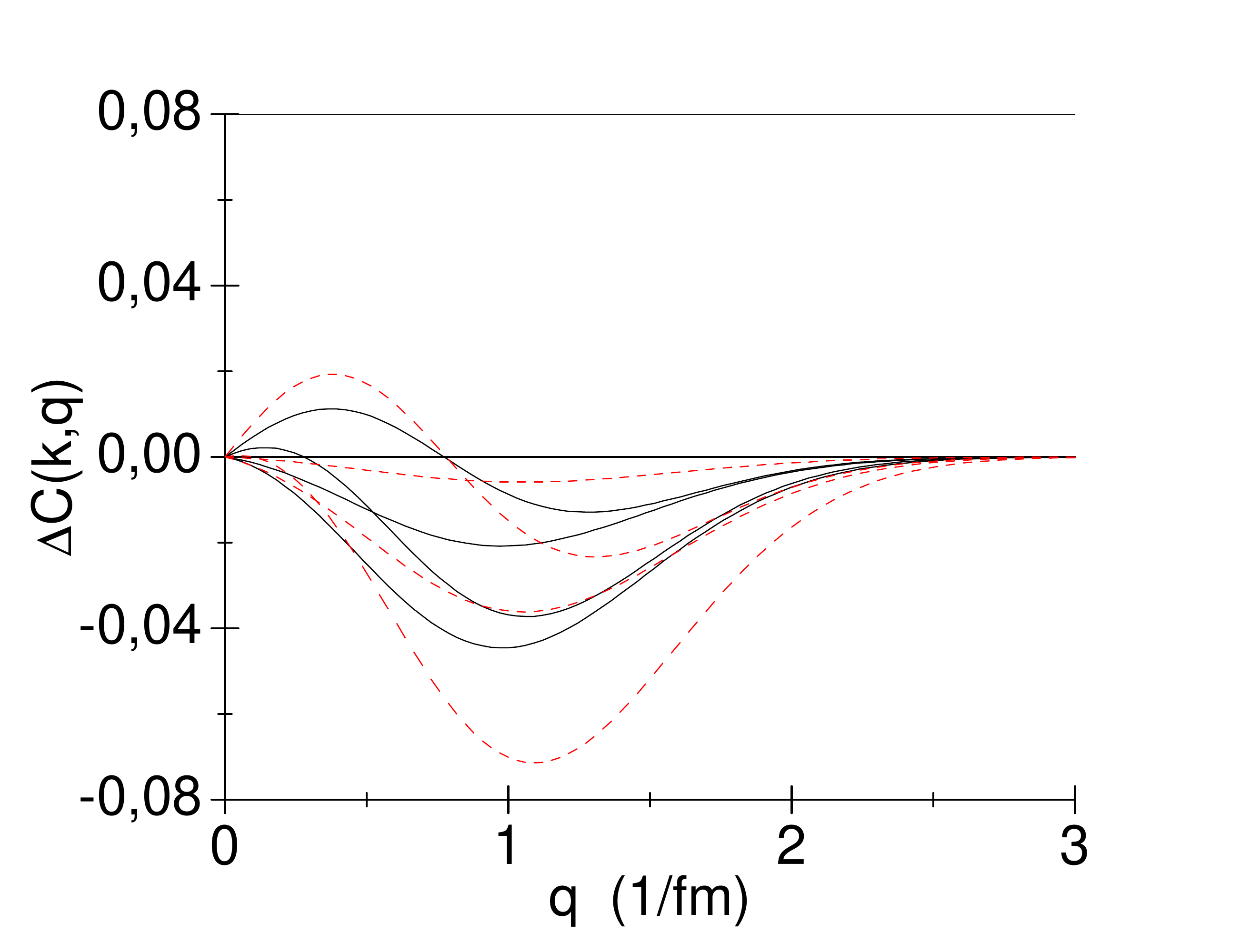}
\end{center}
\vskip -7mm
\caption{ (color online)
Correlation function difference for the weight factors:
$\omega_a = 1.25$, $\omega_b = 0.75$,
$\omega_c = \omega_d = 1.00$
for sources placed at $+x$, $-x$, $+z$ and $-z$ respectively. 
This weight distribution corresponds to the configuration
shown in Fig. \ref{F-16}.
The {\it solid black lines} are for the velocity $v_z=0.5$\,c,
and the {\it dashed red lines} are for the velocity $v_z=0.7$\,c.
Displacement is $d_x = d_z = 1.0$ fm, $T_s = 0.139$ GeV 
and $a = 4b = 4/\sqrt{17}$.
The values of $k$ are: for the {\it solid black lines} 
0.10, 1.00, 3.00, and 5.00 fm$^{-1}$ and
for the {\it dashed red lines}
0.10, 1.00, 3.00, and 5.00 fm$^{-1}$.
For a detector with narrow pseudorapidity acceptance
we could use a value $a > b$.  For this figure 
we have used $a=4b$.
We can compare this with Fig. \ref{F-17},
and we see that for $q > 1.5$ fm$^{-1}$ they are very similar. 
The DCF has a smaller amplitude in the positive direction
for values around $q < 1.0$ fm$^{-1} $, but a 
larger amplitude in the negative direction for values 
around $ 0.75$ fm$^{-1}\, < q < 1.5$ fm$^{-1}$. We also see 
that the DCF is not close to zero for values around 
$2.0$ fm$^{-1} < k < 5.0$ fm$^{-1}$.
}
\label{F-18}
\end{figure}

and it is not dependent on $k$, so for the values 
used in Fig. \ref{F-15} and $q = x\,$fm$^{-1}$ we have
\be
\cosh\left(\frac{0.197}{0.139}\frac{v_z/c}{\sqrt{1-v^2_z/c^2}}
\frac{x}{\sqrt{2}}\right) 
\cos\left(\frac{x}{\sqrt{2}}\right) = 1 \ .
\label{zpeq}
\ee
Since the cosine term must be positive, there are no zero points 
for $\pi /\sqrt{2} \leq x \leq 3\pi /\sqrt{2}$ or $2.23 \leq x \leq 6.66$.

For $x > 3\pi /\sqrt{2}$ the $\exp(-R^2x^2)$ term will be small 
and there would be no correlation difference. 
So we will look at the values $0 < x < \pi /\sqrt{2}$ 

If $v_z$ is smaller than $0.57c$ then there will be 
no zero point for $0 < x < \pi /\sqrt{2}$. For $v_z$ larger 
than $0.58c$ there will be 1 zero point for  $0 < x < \pi /\sqrt{2}$.

The zero point for a given velocity can be found 
by solving equation (\ref{zpeq}) numerically. 
For velocities $v = 0.6c$ and $v = 0.7c$ we have zero points at
$q = 0.83 \, fm^{-1}$ and $q = 1.69 \, fm^{-1}$ respectively.

This indicates the sensitivity of the method and the possibility
to influence it by the choice of the detector directions (via
the choice of $a$ and $b$).

%%%%%%%%%%%%%%%%%%%%%%%%%
\subsection{Emission from four sources}\label{Ef4S}
%%%%%%%%%%%%%%%%%%%%%%%%%

With four sources we can illustrate the possibilities of 
differential HBT method studies in different directions.
The correlation functions can be calculated in general 
for four sources and two detector positions.  This can then
be applied to different detector configurations.

The out component of the four source correlation function with 
weight factors $\omega_a$, $\omega_b$, $\omega_c$, $\omega_d$ 
is given by 

\begin{widetext}  %--------------------------------------------------------

\be
\begin{split}
& C(k_{(\pm)},q_{out})  = 1 +  \exp(-R^2 q^2)\, 
\left[  2\omega_a \omega_b + 2\omega_c \omega_d + \right. \\ 
& \left.
\omega_a^2 \exp\left(\pm \frac{2 \gamma v_z bk}{T_s} \right)
\exp\left(\pm\frac{ \gamma v_z bq}{T_s} \right)\cos(2 a d_x q)+ 
\omega_b^2 \exp\left(\mp \frac{2 \gamma v_z bk}{T_s}\right)
\exp\left(\mp\frac{ \gamma v_z bq}{T_s} \right)\cos(2 a d_x q) + \right. \\
& \left.
\omega_c^2 \exp\left(\frac{2\gamma v_x ak}{T_s} \right)
\exp\left(\frac{\gamma v_x aq}{T_s} \right)\cos(2 b d_z q)+ 
\omega_d^2 \exp\left(- \frac{2\gamma v_x ak}{T_s}\right)
\exp\left(-\frac{\gamma v_x aq}{T_s}\right)\cos(2 b d_z q) + \right.\\ 
& \left.
2\omega_a \omega_c 
\exp\left(\pm \frac{\gamma v_z bk}{T_s} \right) 
\exp\left(\pm \frac{\gamma v_z bq}{2T_s} \right)
\exp\left(\frac{\gamma v_x ak}{T_s} \right)
\exp\left(\frac{\gamma v_x aq}{2T_s} \right)
\cos((a d_x \pm b d_z) q) + \right. \\ 
& \left.
2\omega_b \omega_d 
\exp\left(\mp \frac{\gamma v_z bk}{T_s} \right) 
\exp\left(\mp \frac{\gamma v_z bq}{2T_s} \right)
\exp\left(-\frac{\gamma v_x ak}{T_s} \right)
\exp\left(-\frac{\gamma v_x aq}{2T_s} \right)
\cos((a d_x \pm b d_z) q) + \right. \\ 
& \left.
2\omega_a \omega_d 
\exp\left(\pm \frac{\gamma v_z bk}{T_s} \right) 
\exp\left(\pm \frac{\gamma v_z bq}{2T_s} \right)
\exp\left(-\frac{\gamma v_x ak}{T_s} \right)
\exp\left(-\frac{\gamma v_x aq}{2T_s} \right)  
\cos((a d_x \mp b d_z) q) + \right.  \\ 
& \left.
2\omega_b \omega_c 
\exp\left(\mp \frac{\gamma v_z bk}{T_s} \right) 
\exp\left(\mp \frac{\gamma v_z bq}{2T_s} \right)
\exp\left(\frac{\gamma v_x ak}{T_s} \right)
\exp\left(\frac{\gamma v_x aq}{2T_s} \right)
\cos((a d_x \mp b d_z) q)  \right] \times \\
& 
\left[
\omega_a \exp\left(\pm \frac{\gamma v_z bk}{T_s}\right) + 
\omega_b \exp\left(\mp \frac{ \gamma v_z bk}{T_s}\right) + 
\omega_c \exp\left( \frac{\gamma v_x ak}{T_s} \right) + 
\omega_d \exp\left(-\frac{ \gamma v_x ak}{T_s}\right) 
\right]^{-2} \ .
\end{split}
\label{CFD4eq}
\ee

\end{widetext} %--------------------------------------------------------

\begin{figure}[h] %%%%%%%%%%
\begin{minipage}{.48\textwidth}
\begin{center}
      \includegraphics[width=6cm]{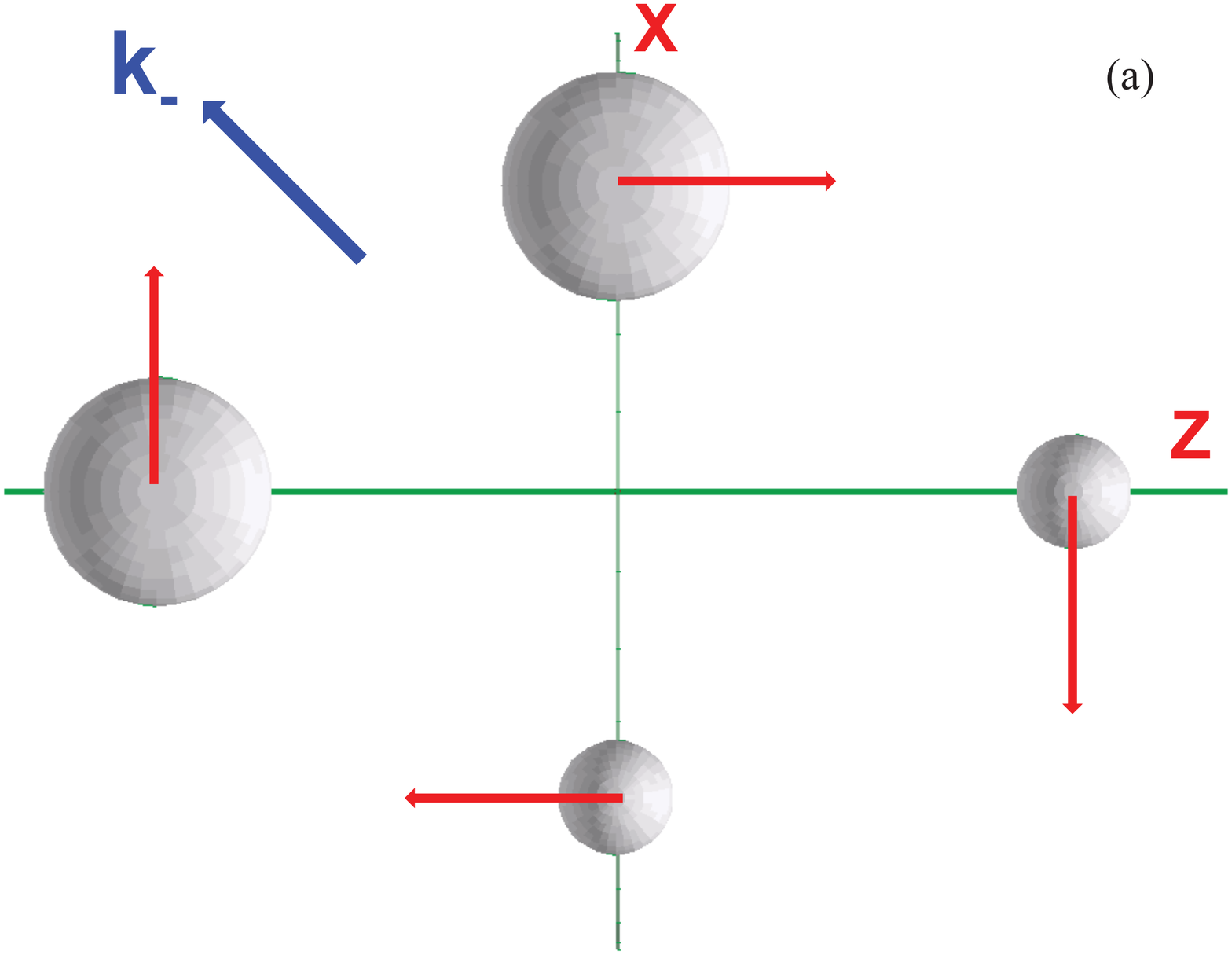}
\end{center}
\end{minipage}
\vskip 3mm
\begin{minipage}{.48\textwidth}
\begin{center}
      \includegraphics[width=6cm]{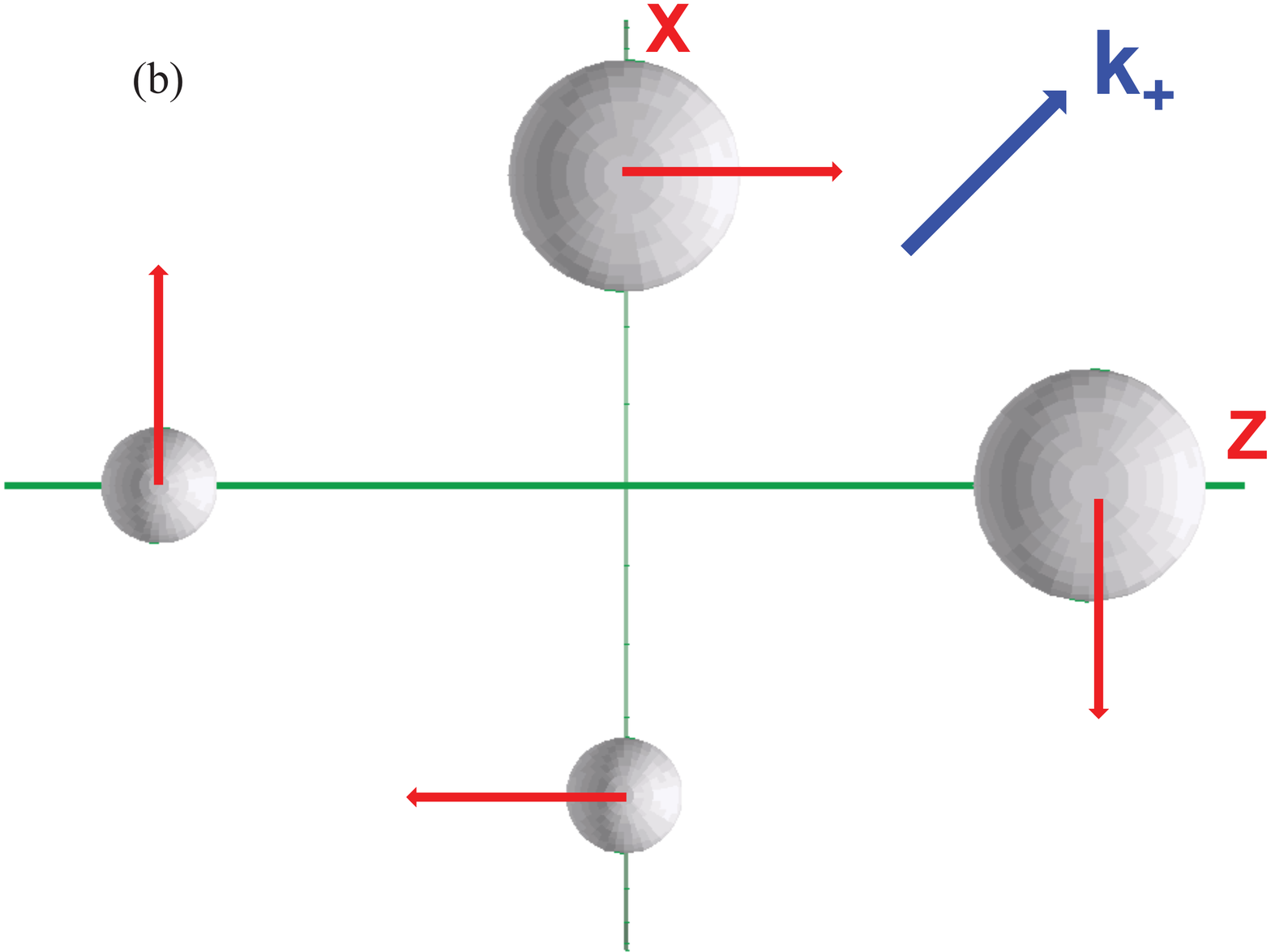}
\end{center}
\end{minipage}
\vskip -4mm
\centering
\caption{ (color online)
Four moving sources in the reaction ($[x-z]$) plane,
separated in the $x-$ and $z-$ directions. 
The sources are moving in the
directions indicated by the (red) arrows. The "tilted" detector
directions are indicated by the (blue) arrows. In the two configurations,
(a) and (b) the detector directions are different and the weights of the
sources are also different, so that the sources closer to the detector
direction have larger weights. }
\label{F-19}
\end{figure}

Two examples on different detector configurations are given in 
Figs. \ref{F-16} and \ref{F-19}. 
We use the same equations as in the two 
source model, Eqs. (\ref{kcomp}) and (\ref{qcomp}).

A source with a larger weight factor is closer to the detector, so that
 $\omega_a$, $\omega_b$, $\omega_c$, $\omega_d$ 
correspond to 
$\vec x_s \equiv (r_x, r_z) = (d_x, 0), (-d_x, 0), (0, d_z), (0, -d_z)$
respectively. 

In case if the detector has a wide pseudorapidity acceptance, then
$\vec k_\pm$ can deviate significantly from
$k_x$, i.e. $b \ge a$ and then the weights 
are maximal for the two sources closest to $\vec k_+$ or $\vec k_-$
as indicated in Fig. \ref{F-19}.

Eq. (\ref{CFD4eq}) can be used to find the difference of the forward 
and backward shifted correlation function.
We will use that $d_x = d_z$, $v_x = v_z$ and $a = b$.

\begin{figure}[b] %%%%%%%%%%
\begin{center}
      \includegraphics[width=7.9cm]{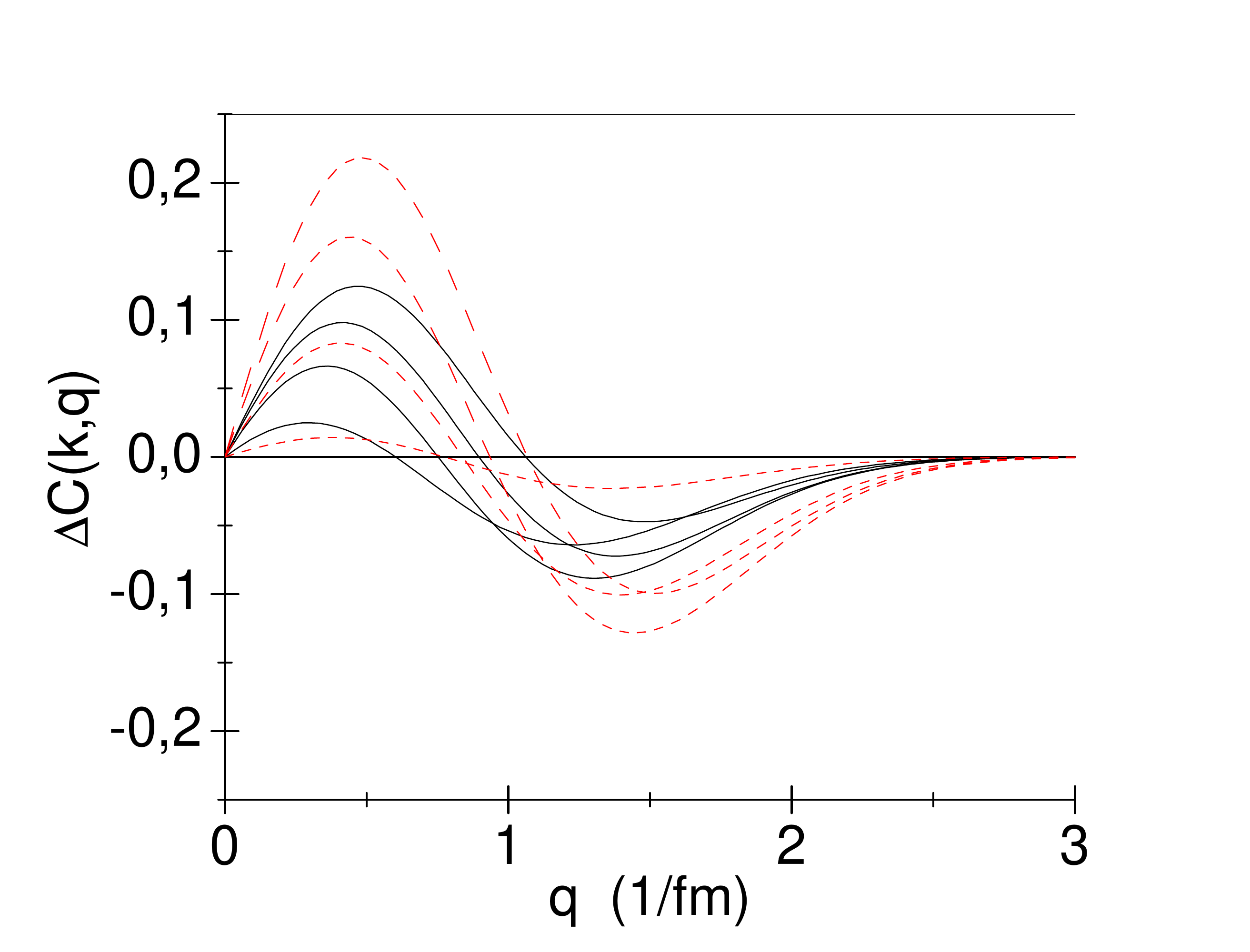}
\end{center}
\vskip -8mm
\caption{ (color online)
Correlation function difference for the weight factors:
$\omega_a = 1.25$, $\omega_b = 0.75$,
$\omega_c = 0.50$, $\omega_d = 1.50$
for sources placed at $+x$, $-x$, $+z$ and $-z$ respectively 
(Fig. \ref{F-19}a)
and 
$\omega_a = 1.25$, $\omega_b = 0.75$,
$\omega_c = 1.50$, $\omega_d = 0.50$
for sources placed at $+x$, $-x$, $+z$ and $-z$ respectively
(Fig. \ref{F-19}b).
The {\it solid black lines} are for the velocity $v_z=0.5$\,c,
and the {\it dashed red lines} are for the velocity $v_z=0.7$\,c.
Displacements are $d_x = d_z = 1.0$ fm, $T_s = 0.139$ GeV 
and $a = b = 1/\sqrt{2}$.
The values of $k$ for both series of lines are
0.10, 0.50, 1.00, and 2.00 fm$^{-1}$.
The difference is larger for smaller values of $k$.  }
\label{F-20}
\end{figure}

Some examples for the differential correlation functions are shown
in Figs. \ref{F-17} and \ref{F-20}. Here due to the simplified few source
model we specified the weight distribution among the sources in a simplified
way. For realistic high resolution fluid dynamical model calculatios
the realistic evaluation of emission probabilities is necessary. 

We can compare Fig. \ref{F-17} with the previously shown
two source model, Fig. \ref{F-15}, and
we see that the amplitudes are similar but the shapes are different.
First of all the sensitivity on the direction of rotation 
remained the same as in the simpler two source model. The two
extra sources, $c$ and $d$ lead to higher amplitude for the 
Differential Correlation Function, while the regular positions
of the locations of the zero points are varying due to more
sources with different weight parameters. 

In Fig. \ref{F-18} we show the DCF for a configuration
where the deviation between $\vec k_+$ and $\vec k_-$
is smaller, like shown in Fig. \ref{F-16}.
This configuration can be applied in detectors where the
pseudorapidity acceptance range of the detector is not wide.
Still the rotation is well detectable. In this configuration
the accurate determination of the reaction plane and the
participant center of mass momentum is more important.

If the detector acceptance is wider, then the two detectors
can be placed at more different angles. This configuration
makes the forward and backward placed sources more
accessible to the forward and backward detectors, respectively.
This is taken into account in the emission weights of our
sources. These weights are now different for the two components
of the DCF!

The result shows the tendency that the DCF
has a similar structure in the 
two source model and the four source model in a resembling
configuration.
Fig. \ref{F-20} has the same shape as Fig.  \ref{F-17}, 
but the amplitude is larger.

For a set of large number of sources, forming a system with 
close to perfect rotational symmetry, a single correlation function
would not depend on the (polar) angle of the detection, and the  
DCF would vanish. Thus, the DHBT method would not be applicable 
for highly symmetric systems, like for a rotating star observed
from within the plane of the rotation.
At the same time for a rotating binary star system the DHBT method
would work. Also the weighting of the sources should be different:
If the observer is in the plane of rotation, the distant
star is shadowed by the front one at some periods, just like emission
from a highly opaque plasma (evidenced by jet quenching). If the
observer is slightly out of the plane of rotation then the two
stars are visible all the time and then the (time dependent)
correlation function would change between the configurations of 
Figs. 4 and 7. This also illustrates the role of symmetric and asymmetric
weightings.

The rotating and expanding final state of a relativistic
heavy ion reaction is of course does not look like a perfect 
wheel, so the four source model is a more adequate approximation
than a wheel would be.

%%%%%%%%%%%%%%%%%%%%%%%%%%%%%%%%%%%
\section{Conclusions}
%%%%%%%%%%%%%%%%%%%%%%%%%%%%%%%%%%%

In this work we attempted to study the possibility of detecting
and evaluation the rotation of a source by the specific use of
the Hanbury-Brown and Twiss method for rotating systems.
Our primary interest was the application for ultra-relativistic
heavy ins where in peripheral collisions at ultra-relativistic 
energies the system can gain large angular momentum.
Nevertheless, some of the conclusions can be applied to macroscopic
systems also, like for past rotating stars.

We selected one of the several methods to evaluate two particle
correlations, which was suitable to study collective fluid systems
with significant and well defined internal fluid dynamical motion.
The obtained standard correlation functions were showing the
consequences of the flow, but for highly symmetric sources
the correlation functions gave symmetric results, which were
invariant for the change of the direction of rotation.

It turned out that it is important to take into account that
the particles reaching the detector cannot reach it with equal
probability from the near side and the far side of the emitting
object. With this fact considered we could obtain correlation
functions, which reflected the properties and also the direction
of the flow.  These results can be used rather generally.

The obtained results have shown that the correlation function
is most sensitive to the rotation if it is measured in the
beam direction (or close to it). This, unfortunately, is
not possible in most heavy ion accelerator experiments, so 
we introduced and investigated a Differential Hanbury Brown and Twiss
method, which made it possible to trace down the rotation
in relativistic heavy ion collisions by measuring the
correlation functions in the reaction plane at nearly transverse 
angles to the beam direction. 
The method is promising and can be performed in most
heavy ion experiments without difficulties, as well as it can be
implemented in different reaction models, like fluid dynamical
models, microscopic transport models and hybrid models.
In full scale theoretical models, the emission probabilities 
from the FO layer have to be considered.
From the general formulas derived in the beginning of the paper
apparently these dependencies can be factorized.

To complement these studies we also applied the method to a high resolution,
3+1D, computational fluid dynamics model \cite{CVW2013}, 
which was used earlier to predict roation, KHI, flow vorticity, and
polarization \cite{hydro1,hydro2,CMW12,BCW2013}. The result shows that
the method can detect rotation, while the effects of
irregular shape, sperical flow, and  specific flow patterns, require
a more extended analysis to separate all these effects form one another.

%%%%%%%%%%%%%%%%%%%%%%%%%%%%%%%%%%%

\begin{acknowledgements}
Enlightening discussions with 
Marcus Bleicher, Tam\'as Cs\"org\H o, Dariusz Miskowiec, 
Horst St\"ocker, Dujuan Wang, 
and scientists of the Frankfurt Institute for 
Advanced Studies are gratefully acknowledged.
\end{acknowledgements}

%%%%%%%%%%%%%%%%%%%%%%%%%%%%%%%%%%%

%%%%%%%%%%%%%%%%%%%%%%%%%%%%%%%%%%%%%%%

\end{document}